# A Survey on Software Defined Networking: Architecture for Next Generation Network

Sanjeev Singh[1] · Rakesh Kumar Jha[1]

**Abstract** The evolution of software defined networking (SDN) has played a significant role in the development of next-generation networks (NGN). SDN as a programmable network having "service provisioning on the fly" has induced a keen interest both in academic world and industry. In this article, a comprehensive survey is presented on SDN advancement over conventional network. The paper covers historical evolution in relation to SDN, functional architecture of the SDN and its related technologies, and OpenFlow standards/protocols, including the basic concept of interfacing of OpenFlow with network elements (NEs) such as optical switches. In addition a selective architecture survey has been conducted. Our proposed architecture on software defined heterogeneous network, points towards new technology enabling the opening of new vistas in the domain of network technology, which will facilitate in handling of huge internet traffic and helps infrastructure and service providers to customize their resources dynamically. Besides, current research projects and various activities as being carried out to standardize SDN as NGN by different standard development organizations (SODs) have been duly elaborated to judge how this technology moves towards standardization.

✉ Rakesh Kumar Jha
  jharakesh.45@gmail.com

  Sanjeev Singh
  sanjeevsinghtara@gmail.com

[1] Department of Electronics and Communication Engineering, Shri Mata Vaishno Devi University, Katra, Jammu and Kashmir, India

# 1 Introduction

The increasing use of multimedia contents and the rising demand for big data analysis require higher networking connecting speeds fulfill the social needs of the world's growing population. It is projected that annual global internet traffic may grow by compound annual growth rate (CAGR) of 23 % from 2014 to 2019, and it may exceed to 1.1 ZB per year or 88.4 EB (one billion GB) per month by 2016 and it is expected to grow 2.0 ZB per year or 168.0 EB per month by 2019 [1]. The increase in mobile connected devices is also being predicted to surpass the number of people on the earth by 2014 and is expected to be 1.4 devices per capita by 2018 [2, 3]. It is also estimated that traffic from wireless and mobile devices will increase manifold in future. This change in scenario of Internet traffic indicates that there will be increase in wireless traffic comprising Wi-Fi, mobile devices by 2019 to 66 % and that of wired traffic will decrease from 54 % in 2014 to 33 % in 2019 [1]. However, it may also be pointed out that by 2019 the Internet connections are estimated to be three times higher than that of global population and per capita Internet traffic will increases from 8 to 22 GB [1]. In India, it is estimated that at present about 980 million mobile users and about 300 million internet connections are operating [4]. The future endeavor will be to cover the remaining population through Digital India Mission. The existing and emerging trend in information and communication technology (ICT) is towards high performance applications, which include mobile computing, ultra high definition (UHD) video on demand, internet of things, cloud computing, fog computing and big data etc. may govern this traffic growth, which can only be controlled by high-capacity dense wave division multiplexing i.e., DWDM circuit switched optical networks [3]. To handle such a huge traffic as well as future Internet applications with efficient and economical delivery of packets is a big challenge for the network administrator. SDN has emerged as a well-organized networking technology in this fast changing scenario of networking, which is accomplished by providing the support to the dynamic characteristics of future network (FN) applications, while having less capital and operating cost through easy to control hardware and simplified software management [3].

SDN has three defining characteristics. First, the ability to decouple the data plane (i.e., forward packets as per the decision taken by the control layer) from the control plane (i.e., routing decision or which analyze the received packet and govern the decision in what way to handle the traffic in routers and switches). Second, SDN provides a unified control plane, in such a manner that a multiple data-plane elements can be controlled via a single software program. The SDN control plane extends direct control over network's data plane elements i.e., switches and interfaces control and data plane via OpenFlow, which is most commonly, used application programming interface (API). Third, this archetype provides networking administrator, a worldwide view of entire network and allows making changes globally instead of making changes on each individual hardware unit (device-centric configuration). This innovative technology and concept was originally proposed by Nicira Networks, which was based on their previous development at UCB,

Standford, CMU, Princwton [5, 6]. Recent work in the field of SDN explores application and extension to a wide range of networks which may include home networks, cellular core networks, enterprise networks, cellular, and Wi-Fi radio access networks etc.

In this article study has been carried out on the SDN literature elaborating its basic concept and architectural principle, indicating recent and future advancement in SDN. We also presented our proposed architecture. Due attention is also given on current researches being carried out. Accordingly, present article is organized as follows: in Sect. 2, we begin with the comparison to understand SDN as advancement over conventional network. Section 3, includes discussion on the motivation behind SDN for adopting it as a FN. Section 4, explains historical evolution in relation to SDN over the past 20 years. Section 5, provide a detail information about SDN technology and its three layer architecture with various techniques used to interface NEs with OpenFlow based SDN, which includes south and northbound APIs, east and westbound APIs. Section 6, covers the domain of Openflow and its advancement with the passage of time. Section 7, illustrate the working of SDN. Section 8, in this discussion is on selective SDN architecture design's applications, technique used to interface NEs with centralized unified controller and their performance. Section 9, discusses in detail our proposed SDHN as FN architecture. Section 10, in this discussion is on SDN's current research projects, indicating its progress towards standardization as NGN. Finally, Sect. 11 concludes with a discussion on "SDN: Architecture for NGN".

## 2 SDN as an Advancement over Conventional Network

Conventional networks implement various dedicated algorithms and set of rules on hardware components like application specific integrated circuits (ASICs) to monitor and control the flow of data in the network, supervising routing paths and responsible for configuring various NEs with each other in the network path [5, 6]. When the packets are received by the routing devices, in a conventional network, it employs a set of rules, which are already entrenched in its firmware to detect the routing path for that packets as well as address of the destination device in the network. Generally data packets are handled in similar manner, which may be directed to the same destination and all this occurs in an inexpensive routing device. Moreover, special routing device i.e., Cisco router may have the ability to treat different packets depending on their nature and contents. It allows the administrator to mark out priorities of different flows through customized local router programming. Thus, the queue size in each router can manage packets flow directly. Such a customized local router setup allows the operators to handle traffic more efficiently in terms of congestion and prioritization control. The current network devices have the limitation on network performance due to high network traffic, which hinders the network performance in terms of speed, scalability, security, and reliability. The current network devices lack the dynamism in operation, which is related to different types of packets and their contents. It may be attributed to inability to reprograming of the network operation due to the underlying hardwired implementation of routing rules and various protocols [5, 7]. To overcome this, suitable handling of data rules are required in the form of software module. It will help in improving control over the network traffic by efficient utilization of network resources, which may lead to a state-of-the-art technology, known as SDN [8]. It also enables a cloud user to use cloud resources such as storage, processing (compute), bandwidth, and virtual machines (VMs) or conduct scientific experiments by creating virtual flow slices more efficiently. The goal of SDN is to provide a framework with open, user-controlled management for the forwarding devices in a network. In it, depending upon the scale of the network, the control plane may have one or multiple controllers. In case of multiple controller environments, a high speed, reliable distributed network control can be formed with peer-to-peer (P2P) configuration. In large-scale, high speed computing network, segregation of data plane from control plane plays an important role in SDN, wherein, switches use flow table for packet forwarding in data plane. Flow table comprise list of flow entries and each entry has three fields i.e., matching, counter and instruction. It leads to improved performance of network in relation to data handling, control and network management. It is due to the fact, that software module (applications) helps administrator to control data flow along with desired change in the characteristics of switching and routing device in network from central location without dealing with each device individually in the network [5]. The comparison between conventional network and SDN is shown in Table 1.

It may also, be stated that an advancement in SDN is to stay as an extra-ordinary evolutionary step, wherein, the OpenFlow standards are also employed along with new services by leveraging virtualization in particular to optical transport network control and management for further improving its capacity domain and efficiency. In view of technological advances of Internet, complex processes are involved and efforts are being made to solve diverse social problems. Research and development are currently underway to realize NGN. An all-optical network is promising technology for FN. In this optical packet and circuit integrated network (OPCInet) offer diverse services, increase functional flexibility along with efficiency in energy consumption with high speed switching in a packet based SDN system in the metro/core network [10].

## 3 Motivation

Motivation for adopting SDN technology as NGN can be visualized from the facts given hereunder:

- To accommodate the fast expending traffic, flow addition investment will be required in the network infrastructure to enhance the capacity of existing computer network. With this, network becomes enormous in size, even for small size organization would require 100-to-1000s of devices. As the nature of networks is heterogeneous, because of the deployment of equipment's, applications and services are provided by different manufactures, vendors and providers, the management of the networks is very complex. Even human factor also contributes to network downtime (faulty) due to manual configuration of the network equipment (NE) as well as network devices outage may also be responsible. Due to these difficulties traditional approach for configuration, optimization and troubleshooting would become inefficient and in some cases insufficient. To overcome these aforementioned problems SDN is touted to provide promising solution by segregating the control logic from the data plane and allow flexibility, efficiency in operation and management of the network via software programs.

Table 1 Comparison between conventional and SDN [9]

| Characteristics | Conventional networking | SDN |
|---|---|---|
| Features | In this, the data and control plane are customized in each node as shown in Fig. 1a. For each problem a new protocol is proposed with complex network control | Segregates the data plane from the control plane as shown in Fig. 1b, with centralized programmable controller that makes the network control simple |
| Configuration | In this, when new equipment's are added into the existing network, because of heterogeneity in network devices manufactures and configuration interface requires certain level of manual configuration procedures, which is tedious and error prone | Unification of the control plane over all kinds of network devices including routers, switches, and load balancers permits automated configuration with centralized validation via software controlling. As such, an entire network can be programmatically configured and dynamically optimized based on network status |
| Performance | Due to heterogeneity among networking devices and coexistence of various technologies, the optimizing performance of the network as a whole is difficult | Provides an opportunity to improve network performance globally with centralized control and having feedback mechanism to exchange information among different tiers of networking architecture |
| Innovation | Unfortunately, in conventional networks certain difficulties are encountered while implementing new ideas and design due to widely used proprietary hardware, which prevents modification for experimentation and adoption | Comparatively, SDN encourages implementation of new ideas, applications and new revenue earning services conveniently and flexibly through programmable network platform |
| Cost | In this, the switching devices both the data and control plane are embedded on the same switch, therefore making switch more complex and costlier | In this, the data plane and control plane are decoupled from each other making the structure of switches simpler and easier to manufacture, which in turn leads to a low cost solution |

- Scalability, reliability, and network performance are main concerns for efficient operation of software defined optical network (SDON) especially at the initial stage when control logic is off-loaded from the switching node. From the study on large emulated network with 100,000 endpoints and 256 switches it is observed that at least 50,000 new flow requests per second are managed by various OpenFlow controller implementations like NOX-MT, Maestro, Beacon etc. [11]. This indicates that surprising large number of new flow requests can be handled by single controller. Thus, the problems of scalability, reliability, and network performance are addressed efficiently.

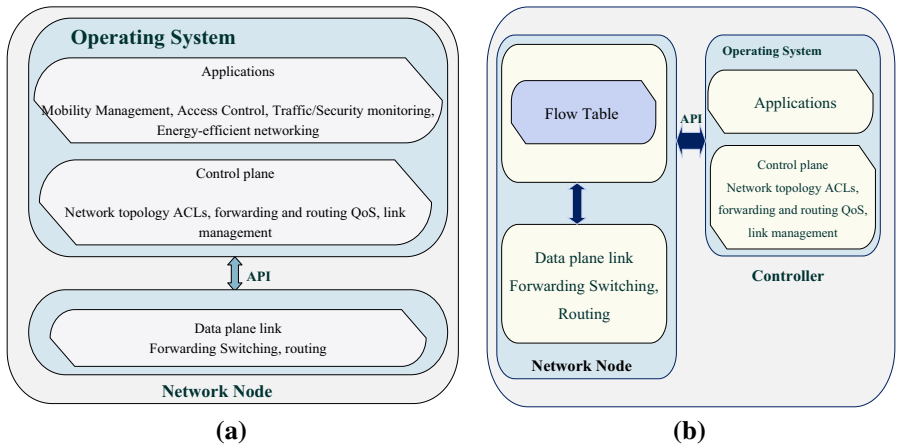

**Fig. 1** Conventional network node compared with the SDN node [9]. **a** conventional approach (each individual network node has its own control and data plane management). **b** SDN approach (the control logic is off-loaded into controller from the network node)

- For on demand mobility and migration of services optical-technology-based SDN can be deployed to simplify implementation of programmable traffic flow control and load balancing arrangements providing inside a data center (DC). Wherein, bandwidth and latency required for different applications (different traffic flow) are taken into consideration [12].
- The current networks are mostly designed for optimum utilization of the underlying infrastructure and the assigned spectrum is overprovisioned. In view of this, new elastic-optical networking (EON) technology was proposed in SDON. Wherein, flexible spectrum bandwidth is allocated to each individual data link without using static wavelength grid. In this flexible bandwidth network, the adaptability is more because the spare spectrum is allocated to re-routed signals, which make it a smart network to utilize its resources with great optimization [13].
- Moreover, SDN can integrate multiple transport technology and multi network domain efficiently and effectively [12].

## 4 Literature Survey on Historical Evolution of SDN

Take off in internet and its historical evolution in relation to SDN is just about 20 years old, which may be divided into various stages as depicted in Fig. 2 and each one has its role to play towards historical evolution. Each event as indicated in Fig. 2 was classified on the basis of working group, author's name, techniques used to interface control-data plane, routing traffic control optimization and operating system. First stage relates to "Active Networks" (period from 1995 to 2001). In this period, programmable functions were introduced and that enabled the network operators to have greater innovations [14, 15]. Second stage relates to "Control and Data plane separation" (period started from 1998 to 2007), wherein, during this period various open interfaces for communication between the control and data planes were developed. In third stage, the "OpenFlow API and network operating systems (NOSs)" (from 2007 to 2014), the extensive use of an open interface and other ways developed made scalability and segregation of control-data plane easy and practicable. During this period, various operating systems were also developed like NOX, Onix and open networking operating system (ONOS). Further, network virtualization (NV) (splicing) played an important role and primarily focused on to find better techniques for route traffic flow and for a wide range of applications [16–18]. However, advancement in NV goes in parallel along with other stages as indicated in the Fig. 2. The brief description to further elaborate the Fig. 2, on the nature of work carried out by various authors is given on Table 2.

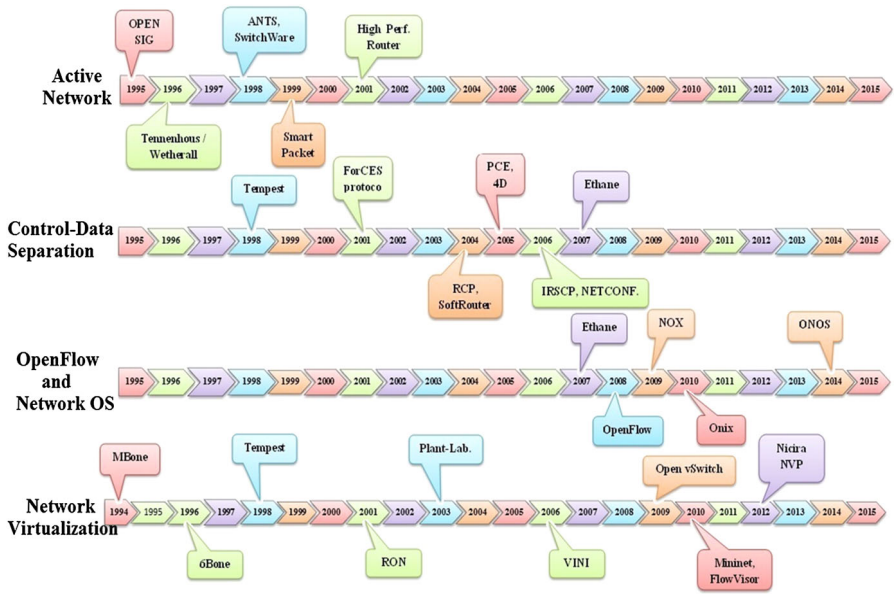

**Fig. 2** Illustrate selective historical evolution in relation to SDN over the past 20 years, with advancement in NV in chronological order [16]

Table 2 elucidate as to how, advancement was brought about in networking from active networking to OpenFlow based SDN and Virtualization, which helped in extending the domain of SDN.

## 5 Literature Survey on SDN Technology and Architecture

SDN a framework to allow network administrators to automatically and dynamically manage and control a large number of network devices, topology, services, packet handling quality of service (QoS) and traffic paths policies using high-level

Table 2 Brief description of selective historical evolution in sdn architecture in chorological order

| References | Stages | Brief description |
|---|---|---|
| [19] | Before active network | Initially, the information carried by the traditional data networks is without any computation i.e., passively transported bits. In 1995, working group on open signalling (*OPENSIG*) conducted sequences of dedicated conferences to make internet, asynchronous transfer mode (ATM) and Mobile networks programmable and more open and extensible |
| [20] | Active network | In this network, each node of the network is inserted with miniature customized program. Moreover, it replaced the passive packets by active "capsules"—(tiny programs) that could be carried in user messages, which are interpreted and executed while traversing at each router/switch |
| [21] | Active network | Earlier, the process of changing network protocol was difficult and lengthy, because there was no automatic mechanism for upgrading multiple protocol functionality. Since internetworking protocol is the basis for interoperability, therefore, an active network toolkit was developed known as *ANTS*. In this mobile code technique was used for the automatic deployment of the protocol at intermediate nodes and end systems |
| [22] | Active Network | In this active network architecture of *Switchware* comprise three layers which includes active packets, active extension and active router to provide flexible, safe and secure performance using cryptography-based security |
| [23] | Active network | *Smart-packets* i.e., user-written network program generates smart-packets, which are further encapsulated into protocol (ANEP) that focuses on reducing un necessary burden on the nodes by proper management and monitoring of the network |
| [24] | Active network | In this for *high performance active router* associated with multi-processor port design is introduced to provide adequate computational means to get increasing demand for higher terabit capacity. Since a single processor system was not sufficient even for single 2.4- or 10-Gb/s link |
| [25] | Control-data plane separation | In *Tempest* (set of components) multi-service network has many different control architecture and demonstrate elegantly the switchlet concept |
| [26] | Control-data plane separation | *Forwarding and control element separation (ForCES)* is proposed by internet engineering task force (IETF). It standardized the communication between the separated control-data plane. In this ForCES NE consisted of multiple forwarding elements (FEs) and multiple control elements (CEs). FE processes the packets as per the CE instruction. To define the protocol between FEs and CEs it used ForCES protocol layer and to transport the protocol layer (PL) message it used ForCES protocol transport mapping layer (ForCES TML) |
| [27] | Control-data plane separation | In each autonomous system (AS) the *routing control platform* (*RCP*) selects routes in lieu of internet protocol (IP) routers (lookup-and-forward switches) for interdomain forwarding packet. This enables simple and reduce error prone traffic engineering. |
| [28] | Control-data plane separation | *Softrouter* architecture segregates the implementation of the control layer task from packet forwarding (data plane) function. It offers increase scalability, reliability, security new functionality and decrease cost |

Table 2 continued

| References | Stages | Brief description |
|---|---|---|
| [29] | Control-data plane separation | In multi-protocol label switching (MPLS) and generalized multi-protocol label switching (GMPLS) networks the *path computation element* (*PCE*) architecture employed in the path computation of label switches individually from packet forwarding |
| [30] | Control-data plane separation | *4D* architecture encompasses four planes that are decision, dissemination, discovery and data. This architecture advocates segregation between routing decision logic and the set of rules that govern the communication between NEs |
| [31] | Control-data plane separation | In *intelligent route service control point* (*IRSCP*) path allocation is done outside the router and acknowledged by external network intelligent. In this, specific focus is given on dynamics connectivity management to optimize the traffic flows across a network |
| [32] | Control-data plane separation | *Ethane* is predecessor to OpenFlow and has a new architecture for enterprise networks to manage policy and security in the network. Focus is given on data-control separation with centralized controller |
| [33] | Control-data plane separation | *NETCONF* stands for network configuration working group established by IETF in 2006, and proposed as a management protocol for altering the configuration of elements in the network. It is originally developed to overcome the shortcoming of the simple network management protocol (SNMP) |
| [32] | OpenFlow and network OS | *Ethane* switch consist of flow-table, a controller (NOX, Maestro, Beacon) and the communication between them is controlled by secured channel. Infact, strong foundation for SDN laid by Ethane |
| [34] | OpenFlow and network OS | *OpenFlow* is proposed by open networking foundation (ONF) to standardize the communication between unified controller and the switches in SDN architecture |
| [35] | OpenFlow and network OS | *NOX* is an "Operating system" and serves as framework which co-ordinate and manage ever evolving technologies. It provides centralized programmable interface evenly distributed for whole network |
| [36] | OpenFlow and network OS | Design and implementation of a platform that fulfill all the requirements of the network is accomplished by the operating system known as *Onix*. It provides distributed control platform to deal with large scale production network on global basis. Control plane transcript with Onix provide global view of the network and use basic state distribution primitives provided by the platform |
| [37] | OpenFlow and network OS | *ONOS* is an open source network operating system that will be available on github. ONOS is a distributed system designed for scale and availability |
| [38] | Network virtualization | *Multicast backbone* (*MBone*) is virtual network that is originated to multicast audio and video. For videoconferencing multicast permits one-to-many and from many-to-many network delivery services |
| [39] | Network virtualization | The *6bone* is established in 1996 by IETF, as a testbed for Internet protocol version 6 (IPv6) and assist transitioning of IPv6 into the Internet. It replaces internet network layer protocols known as internet protocol version 4 (IPv4) |
| [25] | Network virtualization | *Tempest* allows controlling ATM switch at the same time with many controllers by dividing the resources of the switch into switchlet that are controlled by these controllers forming the virtual network |

Table 2 continued

| References | Stages | Brief description |
|---|---|---|
| [40] | Network virtualization | A *resilient overlay network* (*RON*) is an architecture deployed to improve performance by detecting and recovering from path outages with time in a distributed Internet application. It detects the path failure more rapidly than existing inter-domain routing protocols |
| [41] | Network virtualization | *PlanetLab* aims at broad-coverage of network on global basis having a goal to grow 1000 geographically distributed interconnected nodes. In it various applications are run in a slice to evaluate their performance. Infact, a slice acts as a network of VMs, wherein, a cluster of local resources are bound to each individual virtual machine |
| [42] | Network virtualization | *VINI* stands for virtual network infrastructure and used for evaluating the performance of services and protocols. It also provides realistic control over network by deploying software for real routing, along with network events and traffic loads. It helps in running network in slices as has been inferred by deploying PlanetLab in PL-VINI implementation |
| [43] | Network virtualization | VMs are connected to physical interface via *Open vSwitch* rather than directly connected to the network interface cards (NICs) and manage flow of traffic between VIFs adjoined to each other on the same physical host. However, in contrast to physical switches, which are used to connect host with the network, virtual switches are software modules which reside in the host i.e., not present in the physical network |
| [44] | Network virtualization | *Mininet* is an emulator that provides realistic testbed use for design and evaluation the performance of the prototype network architecture and with the help of this exact same tested code can be deployed into a real network |
| [45] | Network virtualization | Flow visor is an evaluation platform that partition the network element by inserting the layer between the control-data plane to avoid the building of separate testbeds, which is expensive to deploy at scale and difficult to maintain |
| [46] | Network virtualization | Distributed virtual network infrastructure (DVNI) offers NV architecture, which addresses the limitation of the existing network methodologies like scalability, dynamic provisioning without restriction, mobility and hardware independence more effectively and efficiently. As a result, DVNI embraces the world's largest virtualized DCs |

languages and APIs. Management includes provisioning, operating, monitoring, optimizing, and managing faults, configuration, accounting, performance and security (FCAPS) using optical media in a multi-domain environment [47]. The block diagram of SDN is as shown in Fig. 3. SDN emphasis on five main features:

- Segregate the data plane from the control plane.
- Obtain global view of the entire network and provide it to the centralized controller.

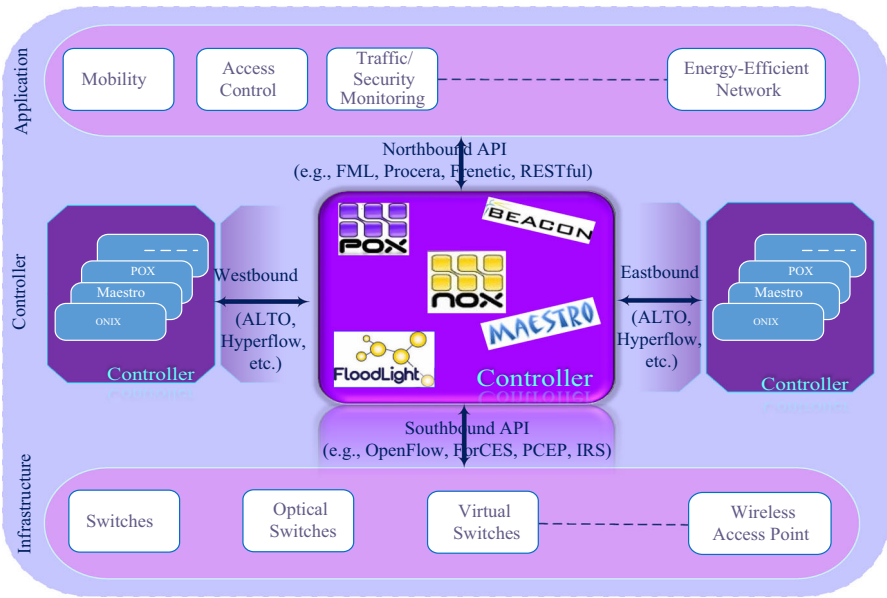

**Fig. 3** Illustrate the functional architecture of SDN, which comprise of Infrastructure, controller and application layer [9]

- Open interfaces between the devices in the data plane and those in the control plane i.e., controllers.
- Network can be programed by external applications.
- Ensure aggregate traffic management.

### 5.1 Infrastructure Layer

The bottom tier of Fig. 3 is known as infrastructure layer. It comprises physical NE like Ethernet switches, routers, optical switches, virtual switches and wireless access point (AP) to name a few and it forms the data plane. All these physical NE's are interconnected to form a single network. The switching devices are interconnected through different transmission media, such as copper wires, wireless radio, and also optical fiber. In this layer, the researcher's interest pertains to efficient operations of switching devices and optimizing utilization of transmission media.

#### 5.1.1 Switching Devices

In a SDN, switching devices simply act as packet forwarding hardware, which is accessible through an open interface, where the control logic and algorithms are off-loaded to a controller. In SDN terminology, these forwarding devices are simply known as "switches". There are two types of switches in an OpenFlow network, such as pure and hybrid. Pure OpenFlow switches completely depend upon a controller for forwarding decisions, whereas hybrid switches support traditional operation as well as protocols and are mostly in commercial use. As in a data plane, these switching devices communicate with the controller to receive the rules, which include packet forwarding rules at a switching level, and link tuning rules at a data-link level and stores the same in its local memory like ternary content addressable memory (TCAM) and static random access memory (SRAM). On the arrival of the packet, these switching device first matches to identify the forwarding rule of the packet and then forward the packet accordingly to next hop. Compared to the legacy networks the packet forwarding rules based on IP or media access control (MAC) addresses, whereas in SDN packet forwarding can also depend on other parameters, like transmission control protocol (TCP) or user data protocol (UDP) port, virtual local area network (VLAN) tag, and ingress switch port [3].

The one major design constraints related to these switches is the efficient utilization of the onboard/local memory. Memory usage depends on the network scale in case of Large scale network huge memory space is required, otherwise constant hardware up-gradation is required to avoid packet dropping or repeatedly directing packet to the controller for further necessary decisions on how to process them and this results in degradation of controller performance [48]. Several solutions are proposed for optimum utilization of the local/onboard memory including route aggregation or summarization and proper cache replacement policy. In this, the memory usage can be reduced by aggregating several routing records with a common routing prefix to a single new routing record having common prefix and with proper cache replacement policy that can improve packet forwarding hit rate. Thus the limited memory can be used effectively and efficiently. Secondly, by improving design of SDN switching devices carefully by integrating various storage technologies to get desired memory size, processing speed and flexibility with reasonable price and complexity. Different storage hardware display varied characteristics, such as SRAM is more flexible being easily scale up, whereas TCAM provides faster searching speed for packet classification, but they are expensive as well as power hungry. Both SRAM and TCAM are used to balance the trade-off between packet classification performance and flexibility [3, 49, 50].

### 5.1.2 Optical Switching

Even today most of the networking equipment that are used in network are still working on the principle of electronic signals, that mean initially optical signals are converted into electrical ones and thereafter these signals are regenerated, amplified or switched, and then again converted back to optical ones. This phenomenon is usually referred as an 'optical-to-electrical-to-optical' (OEO) conversion and with this a significant delay will be generated in the transmission. Optical switches are used to replace the current electronic NEs with optical ones, so that, the necessity of OEO conversions can be eliminated. The benefits of avoiding the OEO conversion stages are significant, as optical switching are inexpensive because there is no need for lots of expensive high-speed electronics.

### 5.1.3 Virtual Switches

These are purposely built for use in virtualized environments and are referred as Open *v*Switch. These switches are used to interface VMs with physical interface via Open vSwitch rather than directly connected to the NICs to manage flow of traffic efficiently. Open *v*Switch is well-matched with almost Linux-based virtualization environments besides QEMU, Xen, KVM, and XenServer [43].

### 5.1.4 Transmission Media

As SDN includes all possible transmission media, such as wired, wireless and optical environments, in order to achieve a ubiquitous coverage, each transmission media have its own unique characteristics, which need specific configuration and management technologies. In order to increase its service area, SDN needs integration with wireless and optical network technologies.

### 5.1.5 Wireless Access Point

It permits wireless devices to have a connection with wired network using Wi-Fi or related standards, where it acts as a central transmitter and receiver of wireless radio signals. Old AP used to support only 20 clients which have now been increased to 255 clients and may further increase with advancement of technology. To increase spectrum utilization in the wireless networks, many advanced technologies have come into operation, which may include software-defined radio (SDR) that permits the control of wireless transmission strategy through software. Due to its similar nature, it can be easily integrated with SDN, wherein, the central controller can manage link association, channel selection, transmission rate and traffic shaping for both clients and APs through the API based on current and historical measurement information, which includes total number of packets, total packet size, and total airtime utilization [51].

### 5.1.6 Optical Fibers

Optical fibers work on the phenomenon of Total internal reflection. As they offer a high capacity with low power consumption, they are widely used in backbones of the network for aggregated traffic management. In optical network reconfigurable optical add drop multiplexers (ROADMs) devices are deployed as the idea of software reconfiguration used in wireless networks, which gives SDN controller a widespread control over all network behaviors including packet forwarding, wireless mode or channel, and optical wavelength [52, 53].

## 5.2 Controller Layer

As shown in the Fig. 3, the middle layer consists of the controllers that are responsible for setting up and tearing down flows and paths in the network. The controller obtains information about capacity and demand required by the networking equipment through which the traffic flows [9]. SDN controller deals with two types of entities, one related to network controlling and the other related to network monitoring. The network controlling includes policies imposed by the application layer and packet forwarding rules for the infrastructure layer. The other is related to network monitoring, in the format of local and global network status. Figure 4 depicts two counter-directional flows in the logical architecture. Wherein, through the downward flow the controller interprets the application policy into packet forwarding rules, in respect to network status. In the upward flow, the controller synchronizes network status collected from the infrastructure for networking decision making. SDN controllers can be segregated into four building components, (1) a high-level language, (2) a rule update process, (3) a network status collection process, and (4) a network status synchronization process [3].

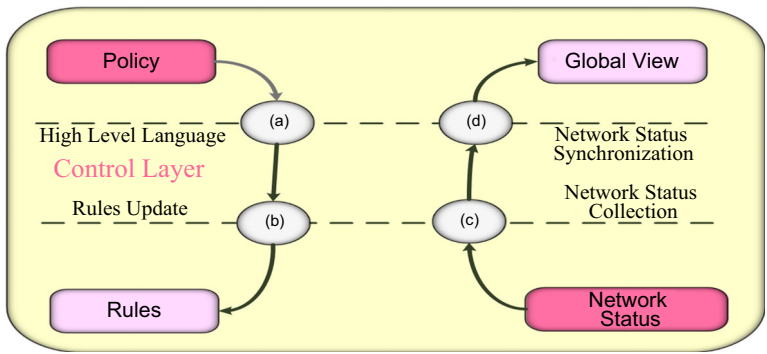

**Fig. 4** Logical architectural design of controller which comprise of four building blocks namely, *a* a high level language for SDN applications to define their network operation policies, *b* a rule update process to install rules generated from those policies, *c* a network status collection process to gather network infrastructure information, and *d* a network status synchronization process to build a global network view using network status collected by each individual controller [3]

### 5.2.1 High Level Language

The key function of the controller is to translate application requirements into packet forwarding rules. This function dictates a communication protocol i.e., a programming language between the control layer and the application layer [3]. Three important characteristics of SDN language are:

1. Network programming language should be capable to offer the resources that can enquire the state of network. The runtime environment of the language has the ability to collect the information about the state of network as well as statistics, and then provide this information to the application layer.
2. The language should have the capability to express policies of the network in relation to the packet forwarding behavior. It should be capable to combine policies of various network applications. It should also be able to resolve the conflicts, if so generated by the network applications.
3. Due to the existence of varied network policies; it is not convenient to reconfigure the network. Therefore, runtime environment of the language should activate the necessary update process of the devices so that it may assure the preservation of access control, avoidance of forwarding loops or other invariants are met. Frenetic, its successor Pyretic, and Procera are the common befitting SDN programming languages that are required to fulfill the present requirements [54].

### 5.2.2 Rules Update

SDN controller is accountable for generating packet forwarding rules. It also describes the guidelines for the packet forwarding and installs them into suitable switching devices for proper operation. At the same time, these forwarding rules in the switching devices are required to be updated due to changes in network configuration and dynamic control, such as directing traffic from one replica to another for dynamical load balancing [55] and network recovery after unexpected failure. Due to the dynamic nature of the SDN the consistency of the rules get updates and reserved to ensure the proper operation of the network, such as, loop free, no black hole, and security. Rule update consistency can be done in different way; however two of them are mentioned hereunder:

- *Strict Consistency* It makes sure that either the original rule set or the updated rule set is used. This consistency is implemented at the level of processing each-packet or in a per-flow level, where all packets of a flow are processed by using either the original or the updated rule set.
- *Eventual Consistency* It makes sure that the upcoming packets use the updated rule set eventually after the update procedure finishes and allows the earlier packets of the same flow to use the original rule set before or during the update procedure [3].

### 5.2.3 Network Status Collection

In this information about network status indicated by traffic statistics, which comprise packet number, duration time, data size and bandwidth share is collected by the controller through upward flow, and accordingly global view of entire network is constructed. This information i.e., network topology graph is to provide the application layer for further necessary decision [56]. In the working of network status collection, each switching device in the network collects and stores the statistics of local traffic in its onboard memory and this information is recovered by the controller via a "pull" mode or by the "push" mode [3].

### 5.2.4 Network Status Synchronization

Assigning control to a single centralized controller may lead to performance bottleneck and to overcome this multiple controllers are deployed in P2P acting as back-up or replicate controller [57]. To ensure proper operation of network, all controllers should have the ability to build and maintain a global network view using network status collected by each individual controller [3]. Tootoonchain and Ganjali [58] place in practice hyperFlow that permits sharing of global view of network among multiple controllers.

### 5.3 Control Layer Performance

In SND networks, performance rely on the control layer, whose performance is further limited by the scalability of the centralized controller. In such network, all the activities such as, on the arrival of first packet of each flow switching devices have to request for packet forwarding rules, collecting information about the network status and rules updates requires continuous communication between the controller and switches, which leads to unnecessary bandwidth consumption and latency of frequent communication and thus affects the control layer performance [3]. To address these aforementioned issues, to increase processing abilities of a single controller in the control layer and to decrease the frequency of request process by the controller following efforts are to be made:

#### 5.3.1 Increasing Processing Ability

As controller is an essential part of the SDN, the conventional techniques such as parallelism and batch processing can be used for improving controller performance on request processing. These are already in use in Maestro [59, 60], NOX-MT [11], and McNettle [61] controllers.

#### 5.3.2 Reducing Request Frequency

As all the transactions in the network are controlled by the controller, this frequent requesting to the controller may result in longer delay in response from the SDN controller side. Many strategies have been adopted to decrease request frequency. Two of them are given here: (1) modify switch devices, so that requests can be handled in or near the data plane. In this approach, Yu et al. [62] suggests that forwarding rules are distributed among each "authority switches" in the data plan, which can handle request and divert each packet through it, which need to access appropriate forwarding rules without requesting to controller for rules. (2) By proper organization of the structure in which switching devices are deployed, can also help in improving the overall performance of the control layer [3].

#### 5.3.3 Performance Benchmarking

In SDN controller performance benchmarking can be used to indicate the performance bottleneck, which is an important parameter to increase the processing ability of controller. The two tools that are designed for measuring controller performance benchmarking are Cbench [63] and OFCbench [64].

### 5.4 Controller Interface

*5.4.1 Protocol Options for the Southbound Interface*

As shown in Fig. 3, link that connects control layer with the physical layer via API is referred as southbound API. OpenFlow is the most commonly used south bound interface. It was standardized by the ONF established in 2011. The main purpose of the OpenFlow is to standardize the communication between the switches and the software-based controller in SDN architecture [65]. Whereas, ForCES can also be used for exchange of information between control and data plane as a second southbound interface option. As compared to OpenFlow, ForCES is having more flexible mode and rich protocol features. But due to some disruptive business model and lack of open source support, it is not so widely adopted as OpenFlow. However, OpenFlow still has to learn more from both merits and shortcoming of ForCES for its future success [3].

In SoftRouter architecture, the control plane functions are segregated from the packet forwarding data plane functions and provide dynamic association between control and forwarding plane elements, which permit the dynamic allocation of control and data plane elements. This architecture has certain advantages over the border gateway protocol (BGP) with regards to its reliability [28]. Both ForCES and SoftRouter have resemblance in their operation with respect to OpenFlow and can be used as alternative southbound interfaces.

One of the most commonly used protocol between the control layer and the physical layer is PCE protocol, a special protocol that permits the path computation client (PCC) to request for path computation from PCE and PCE protocol also acknowledges for the same as shown in the Fig. 5. PCE may have the complete knowledge/picture of flow and path in the network. When a new client comes online, PCC sends request for path computation to the PCE as the PCE have a complete traffic engineering database. The client traffic requirement is calculated and superimposed on the current network's topology. This protocol was developed by IETF PCE working group. Moreover, PCE may be centralized or may be distributed in many or every controller/router [47].

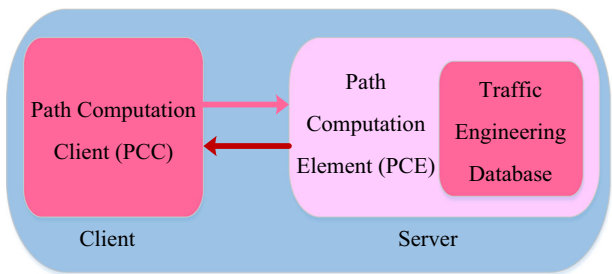

**Fig. 5** Illustrate the working of PCE protocol [47]

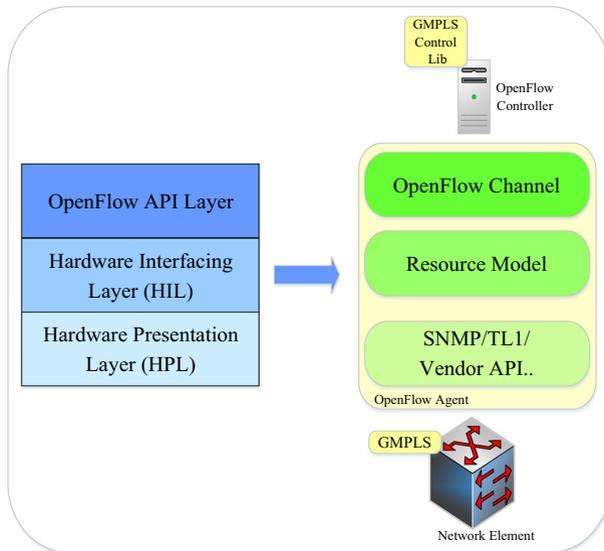

**Fig. 6** OpenFlow agent abstraction [12, 66]

### 5.4.2 OpenFlow Protocol Extension for Optical NEs Interface in OpenFlow-Based Optical Network

OpenFlow-based optical network uses the services of the centralized controller to manipulate the NEs i.e., optical switches in the forwarding data plane via a secured Open-Flow link protocols. In this network, OpenFlow switches are responsible for performing forwarding function according to the flow table entries and the controller host network application, such as path computation, energy management etc. OpenFlow protocols provide the abstract information related to the network, so that the control plane decision can be enforced by inserting flow rules or action into the flow table of OpenFlow switches. Three main messages are generated by the OpenFlow protocol; they are Switch feature, Flow_Mod, and CPort_status [66]. The detail of which is given in Table 3.

**Table 3** Brief description of OpenFlow protocol messages [66]

| S. no. | Messages | Brief description |
|---|---|---|
| 1 | Switch feature | Used by OpenFlow switches (nodes) to describe the capabilities and limitations of NE in the network |
| 2 | Flow_Mod | Used by the controller to add new flow entities and accordingly define flow switching action in each OpenFlow switches/nodes in the network |
| 3 | CPort_status | Gives the ports characteristics changing information e.g. if link is not in operation or down and if the bandwidth of a particular link is updated |

Presently, in the absence of provision of optical equipment vendor's support for OpenFlow [66], the alternative techniques are proposed in literature and some of them have been described in brief in the Table 4.

Table 4 Describe alternative techniques to interface optical devices with OpenFlow-based SDN

| References | Technique/approach | Brief description |
| --- | --- | --- |
| [3, 67, 68] | Virtual ethernet interfaces (*veths*) | Any optical network consists of various NEs namely ROADM, optical switches, wavelength cross-connects (WXC), photonic cross-connect (PXC) etc. OpenFlow-based optical node uses PXC to explain how OpenFlow protocols are used to control optical node as shown in Fig. 10a, b. The combination of OpenFlow switch and PXC is known as OpenFlow-enable PXC (OF-PXC) and this combination is controlled by NOX controller via OpenFlow protocol. In this approach i.e., "veths" was introduced to control this optical node through OpenFlow protocol in the OpenFlow switch. In this i.e., veths approach, OpenFlow switch obtain hardware abstraction information (physical structure of PXC) and provide to NOX controller for efficient control of cross-connection within the PXC with the help of OpenFlow protocol |
| [12, 66] | OpenFlow agent | As in SDN networks the underlying infrastructure/resources information is abstracted and provided to the control layer controller. The abstraction is done to hide the complexity and the technological details of the underlying heterogeneous NEs. Hardware abstraction on vender devices can be done by two ways: (1) hardpath; in this approach, abstraction layer functionalities and flow matching are implemented by using Fast hardware e.g. TCAM. However, keeping in view the fact that in current optical devices hardware abstraction layer is not embedded, therefore, it necessitated the use of Software-based approach i.e., softpath. (2) softpath; in this approach the functionalities of hardware abstraction layer and flow matching is done on the bases of software module called as OpenFlow agent. OpenFlow agent is placed on the optical node, so that it may be able to support OpenFlow protocol. It composes of three layers namely NE's management interface, resource module and OpenFlow channel. *NE's management interface* i.e., SNMP, vender API etc. is used to communicate with the data plane, where direct OpenFlow protocol implementation is not supported and provide the hardware presentation layer (HPL) functionality. *Resource model* to implement the functionality of hardware interface layer (HIL), a generic and novel Resource model was developed and implemented to maintain the NE's configuration (wavelengths, port capabilities and switching controls). *OpenFlow channel* is also included in the OpenFlow agent, this channel provide communication with the OpenFlow controller as show in Fig. 6 |

**Table 4** continued

| References | Technique/approach | Brief description |
|---|---|---|
| [69] | GMPLS | MPLS stand for multi-protocol label switching. In name multi-protocol indicate that it can be used with many different protocols and Label Switching indicates that it is a switching protocol. It is used to encapsulate data packets by adding label to them and provide a predictable path for traffic engineering. Traffic engineering means having full control over the path that packet takes and design the network accordingly. With control over the path selection, traffic can be forced on under-utilized links. MPLS is also referred as layer 2.5 protocol. The architecture of MPLS comprise of data plane for forwarding packets and control plane for Label-Switch path (LSP) establishment i.e., unidirectional packets flow from beginning to end. GMPLS extends the services of the MPLS to support various data transport technologies. GMPLS includes three new interfaces in addition to previous packet-switch capable (PSC) interface. The three new interfaces are named as time-division multiplex (TDM) capable, lambda (wavelength or waveband) switch capable (LSC) and fiber-switch capable (FSC). With GMPLS it is possible to implement unified control plane, that can be supported by broader range of network element with different transport capabilities like ATM switches, IP routers, optical cross-connects (OXC), SONET/SDH cross-connect, PXC etc. With this it is possible to have interoperability in a multi-vender network and provide seamless internetworking connectivity between various types of NEs |
| [66, 70] | Hybrid i.e., combination of OpenFlow agent and GMPLS | OpenFlow-based/SDN optical network utilizes GMPLS, which has the capability of using optical functionalities like power equalization, impairment etc. It also provides applications for control network through path computation and management. However, in spite of these advantages it has failed to encourage the providers due to its inflexible and closed architecture. To overcome this difficulty hybrid i.e., OpenFlow agent and GMPLS control plane approach is advised to take the advantage of GMPLS for its control functionalities and OpenFlow for openness and flexibility by using extended optical OpenFlow i.e., OpenFlow agent. In hybrid GMPLS-OpenFlow technique where NE function acts as an OpenFlow enabled switch |

### 5.4.3 Northbound API for Network Applications

Like southbound interface, the control layer also provides a similar interface with application layer known as Northbound interface to extend services of the application running in the top layer. Service specific application like traffic engineering tells the controller about the path laid for the packets flow from beginning to end, whereas, controller modifies the flow table of the switches with the help of appropriate command. The famous programming languages that are used to write the application programs are flow-based management language (FML), frenetic, pyretic, and procera. FML [71] earlier known by the name flow-based security language (FSL) [72] is a language, which is used in SDN for describing network connectivity policies. Frenetic [73, 74] was introduced to remove complicated asynchronous and event-driven communication between the switching devices and SDN application. Sequential composition was introduced by pyretic [75], which permits superimposition of one rule to another rule while packet processing and it abstract the network topology information that contain maps between physical and virtual switches [3].

### 5.4.4 Interface Between Controllers Operate with East and Westbound APIs

Control plane has two arrangements: one is *physically* and second is *logically* centralized. Physically centralized control plane has a single controller in the control layer and communicate with large number of NEs to collect information of global network view for optimum and intelligent control of the underlying resources e.g. routing protocol design for controlling and managing flow of traffic, with centralized single controller, which may have possibility of failure and potential bottleneck while interacting with large number of NEs. Therefore, single controller deployment is not suitable solution due to lack of scalability and reliability. An alternative approach that is logically centralized control plane offers more scalability and reliability. It consists of physically distributed CEs and each CE is connected to each other through an interface so-called East and Westbound interface [54] as shown in Fig. 3. ALTO stands for application layer traffic optimization used to optimize P2P traffic and developed by IETF working group. Currently, problem with the P2P traffic is if two controllers have more compatibility, a lot of traffic will flow between two of them as compared to others, so ALTO server have the knowledge of all nodes in the network, which helps in defining where they are located, what are their characteristics, how far they are from each other, and what is link bandwidth they have, therefore ALTO provides guidance for peer selection. When ALTO client requests the server for appropriate peers, in response to this a best possible list of potential peers is provided to the client for better communication between them [47]. HyperFlow application that sits on NOX controller and activates with an event propagation system [9]. HyperFlow provide a platform to share synchronized global network view constantly with multiple controllers. It uses a publishing/subscribed system to report whenever a change is sensed in the network status e.g. when a link failure is detected by the controller in its domain, it immediately publishes an event about the change via publish/subscribe system so that other controller may know about the change in network status and with this effect a new updated status is forwarded to each controller [3, 58].

### 5.5 Application Layer

The top tier in the block diagram Fig. 3, that resides over the control layer is called Application layer. SDN applications continuously abstract information about the global network status via south and northbound using protocol like ALTO [76], and eXtensible session protocol (XSP)/eXtensible messaging and presence protocol (XMPP) [77] and manipulates the physical NEs using high level programming languages for writing various functional applications, such as energy-efficient networking, security monitoring, access control link, traffic engineering, PCE etc. The insight detail of the application layer is given as under:

#### 5.5.1 Adaptive Routing

The two main functions that are performed by any network are packet switching and routing. Currently, packet switching and routing design are based on distributed approach, but this approach has certain limitations which may include slow convergence, complex implementation and restricted capability to achieve adaptive control. On the other hand, SDN operates on the principle of closed loop control, wherein, global network status information is constantly fed to the applications so that adaptive control of the network is possible. Two popular adaptive routing applications in the SDN domain are Load balancing and cross-layer design.

*5.5.1.1 Load Balancing* In DCs, most commonly used technique is load balancing to have efficient resource usage. To increase throughput, reduce response time and avoid overloading of network, a front-end load balance is deployed in the DC, so that each request of the clients is directed to a particular server. Allocating a dedicated load balancer is an expensive approach and may create bottleneck as all requests are processed by the same. Wherein, SDN load balancing is done by using various algorithms for packet forwarding rules. Koerner et al. developed and implemented differentiated load balancing algorithm to have control over different types of traffic, such as web traffic and e-mail traffic [3, 78].

*5.5.1.2 Cross-Layer Design* In layered architecture, the cross-layer design is responsible for increasing the integration between various entities that are lying at different layers as in OSI reference model entities at different layers are permitted to exchange information within each other. Since, SDN applications have the capability to access the network status information, this cross-layer design is most suited to deploy for increasing the overall efficiency of the network. Wang et al. introduce a cross layer approach, which has the potential to dynamically configure the underlying network element taking the benefits of high speed and re-configurability of SDN switching devices including optical switches [3, 79].

#### 5.5.2 Boundless Roaming

Mobile devices like tablet and smartphones have the wireless access to the internet, which need continuous connectivity for ubiquitous communication. To achieve un-interrupted services, seamlessness handover has to play a vital role for its applications. However, in the current literature, handover is limited to single carrier network having the same technology. Keeping this in view, SDN provides unified central plane possessing different carriers with varied technologies to enable boundless mobility. Researchers have developed various handover technologies based on SDN.

Yap et al. [80] workout a handover algorithms involving network between WiFi and WiMax, which includes hoolock, to exploit multi-interface on a device and multi-casting. In another study with Odin, where in unique basic service set identification (BSSID) has been allocated to each client connection. In this technique, BSSID of one Physical Wireless AP is swapped with another BSSID of nearby AP during handover, which display low delay in re-association, no throughput degradation and minimum impact on HTTP downloading in either a single or multiple handover [81].

### 5.5.3 Networking Maintenance

In configuration error, which leads to network failure, major contribution is of human factor. Wherein, individual diagnostic tools such as ping, traceroute, tcpdump and NetFlow fail to provide automatic and compressive network maintenance solution. The centralized and automated management techniques inherited in SDN help in reducing the configuration error. Xia et al. [3] introduce a fast restoration technique for SDN in which as soon as, failure of the network is detected, the controller immediately calculates a new path for un-interpreted traffic flow with update packet forwarding rules.

### 5.5.4 Network Security

Currently, for network security firewalls and proxy servers are deployed to protect the physical network breach, but due to heterogeneity in the network, architecture authentic implementation of these techniques is a great challenge for the network operators. Whereas, SDN provides unified centralized control plane, which makes it convenient to implement merge and check policies to prevent security breaches [3].

### 5.5.5 Network Virtualization

In NV, physical network is sliced into multiple virtual network entities and further allotted them to varied user and controllers. However, in SDN, FlowVisor is most commonly used tool which permits slicing of the physical network resources such as topology, flow space (data flow table in switching), bandwidth, switching devices CPU, and control channel to create virtual network for research experimentations [3, 82].

### 5.5.6 Green Network

In this, main concerned are economic and environmental benefits. Heller et al. proposed an energy-aware data link adaption mechanism to work out a minimum data link and switching devices for DC network for energy efficient operations [3, 83].

## 6 OpenFlow

OpenFlow was originally proposed by Nick McKeown in 2008 and was further, standardized by ONF in 2011. OpenFlow was developed to standardize the communication between OpenFlow switch and the software-based controller in SDN architecture. Figure 8 shows advancement in OpenFlow in chronological order and Table 5 illustrates brief description of selective OpenFlow controller. Open-Flow decouples the control plane from the data plane and most commonly used protocol for southbound interface. The architecture of OpenFlow comprises three main components as shown in Fig. 7a; (1) OpenFlow-compliant switches constitute the data plane; (2) the control plane has one or more OpenFlow controllers; (3) the control plan is connected with switches through a secure control channel i.e., OpenFlow interface. An OpenFlow-compliant switch in the data plane simply acts

**Table 5** Brief description of selective OpenFlow controller [54, 65]

| References | Controller | Language | Open source | Comments |
|---|---|---|---|---|
| [35] | NOX | C++/Python | Yes | Designed by Nicira Networks in 2008 at Stanford University and licensed with general public license (GPL), first controller coded in both C++ and Python |
| [90] | POX | Python | Yes | POX cleft from NOX, designed by Nicira Networks in November, 2013 at Stanford University and licensed with apache public license (APL), general purpose controller coded in Python |
| [91] | Maestro | Java | Yes | Network OS written in Java provisioned with multi-thread having developed from Rice University and licensed with lesser general public license (LGPL) |
| [92] | Beacon | Java | Yes | Firstly, originated from Stanford University and licensed with BSD, provisioned with both multi-thread and event-based operations |
| [93] | Floodlight | Java | Yes | Cleft from Beacon, which operates with physical and OpenFlow vSwitches, adequately documented, which makes it designer-friendly. Funded by big switched networks and accessible to apache public license (APL) |
| [94] | Opendaylight | Java | Yes | Introduced by Linux Foundation, issued via eclipse public license (EPL) and OS has no limitation |
| [95] | Flowvisor | C | Yes | Develop by Stanford/Nicira as a special purpose controller implementation |
| [96] | RouteFlow | C++ | Yes | Develop by CPqD as a Special purpose controller implementation |

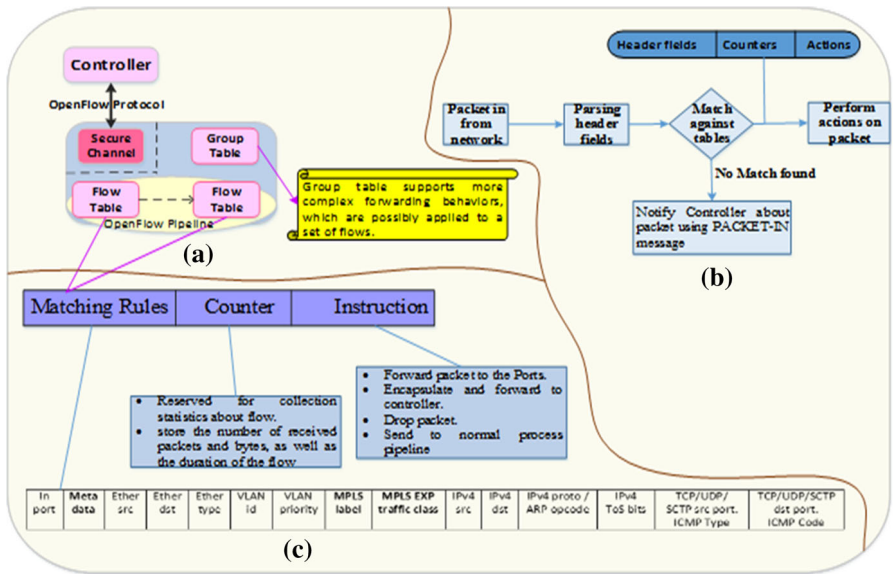

**Fig. 7** Illustrate OpenFlow: **a** architecture of OpenFlow, **b** mechanism for packet forwarding in a switch with OpenFlow, and **c** flow table entry [54]

as device for forwarding packets according to its flow table. A flow table comprises list of flow entries. Each entry has match fields, counters and instructions as illustrated in Fig. 7c. The mechanism for packet forwarding with OpenFlow is illustrated in Fig. 7b. When packet is received by the switch, it analyses the packet header and matching is done with the entries in flow table of the switch. If the flow table entry is matched with the header of the packet, then that particular entry is considered. If several such entries are found, in that case packets are matched on the bases of prioritization i.e., most specific entry will have the highest priority as shown in Fig. 7c. After the matching and selection process is over, and then the counter of the flow table entry is updated. Finally, the switch executes the action on the packet in accordance with the entry in the flow table e.g. forward packets to the port, encapsulate and forward to the controller, drop packet and send to normal processing pipeline. In case, if the packet header does not find match with the flow table entry, then the switch notifies the controller and encapsulates the packet and sends it to the controller with PACKET-IN message as first byte of the packet. On receiving PACKET-IN notification from the switch side, the controller finds the exact action for the packet and installs one or more suitable entries in the requesting switch flow table and then packets are forwarded according to the rules. This is triggered by explicit PACKET-OUT messages. Usually, the controller lays the entire path for the packet by altering the entries in the flow tables of all switches on the path in the network. A software program, called the controller, all the flow table entries are manipulated and populated by the controller by insertion, removal of flow entries and modifications. With this, the controller is able to modify the behavior of the switch with respect to forwarding the packets. To communicate with the switches, the controller uses the secure channel [54].

| Version 1.0 | Version 1.1 | Version 1.2 | Version 1.3 | Version 1.4 |
| --- | --- | --- | --- | --- |
| •Released 31 Dec., 2009. •Single flow table with queues. and each queue is dedicated to a port. •Flow table entry comprise of Header Fields, Counters and Actions. •In this, Match Fields Comprise of **Ingress Port, Ethernet:** src.dst, type, **IPv4:** src, dst, proto, ToS, **TCP/UDP:** src port, dst port | •Released 28 Feb., 2011. •Pipeline of multiple flow table and Group table were introduced. •Due to pipeline new metadata field is required. •Flow table entry actions is replaced by instructions. •Addition in Match Fields over OF 1.0 **Metadata, MPLS:** label, traffic class | •Released Dec., 2011. •IPv6 •OpenFlow compliant Switch may be connected to multiple controllers simultaneously with master/slaves concept for load balancing and fast recovery in case of netwok failure. • Match Fields in addition to previous version i.e., 1.0 and 1.1 **OXM, IPv6:** src,dst, flowlabel, ICMPv6 | •Released 25 June, 2012. • When packets are send from switch to controller cookies along with specific durations fields can be added. • Meter table entry introduce and comprise of Meter Identifier, Meter Bands and Counters. • Multiple controller provision extended. •Provider Backbone Bridge (PBB) added with other protocol small improvement. | •Released 15 Oct., 2013 • Bundles and Synchronized tables along with Optical ports added. • Introduce more flexibility for multi-controller mechanisms along with new codes for error. • Improvement in Eviction and vacancy events as well as in PBB. • Alteration in default TCP port to 6653 •Match Fields in OF 1.3 and 1.4 in addition to previous version i.e. 1.0,1.1 & 1.2 **IPv6 Extension Headers** |

**Fig. 8** Advancement in OpenFlow in chronological order [54, 65, 84–89]

## 7 Working of SDN

In SDN, Control is taken out of the individual network nodes and placed at the separate centralized controller. This controller performs various functions, such as route management, network visibility, network provisioning, NV, and orchestrates network overlays. As shown in the Fig. 3, NOS controls the SDN switches that gather information using the API, which refers as southbound and manipulates forwarding plane by providing an abstract model of the network topology to the SDN controller hosting various applications. The controller can, therefore, use the detail know how of the network for optimizing flow management and support service-user requirements of scalability and flexibility. For example, dynamic allocation of bandwidth can be done into the data plane from the application.

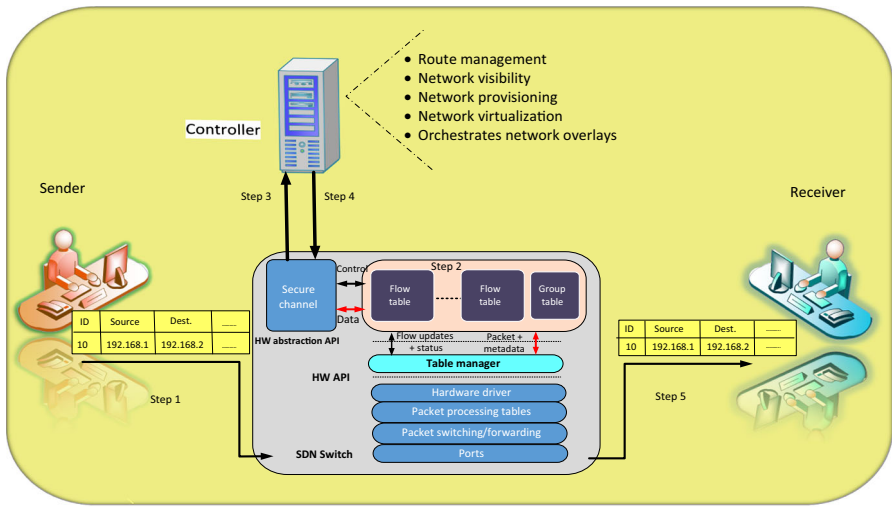

**Fig. 9** Illustrate the operation of SDN controller and switch [9]

As illustrated in the Fig. 9 when packet arrives at the switch from the sender as a first packet of a new flow (Step 1), for this packet SDN switch checks a flow rule by matching it with the flow tables entries in the SDN cache (step 2), if a matching entry is detected in the flow table of the switch, the instructions associated with the specific flow entry are executed e.g. packet/match fields, update counter, metadata and action set. Thereafter, packets are directed towards the concerned receiver (step 5). In case, there is non-availability of the match in the flow table of the switch, then packet is directed to the controller via a secured channel (step 3). Controller analyses the packet for the source and destination IP address and accordingly updates flow table entries of the switches in the path through the southbound API i.e., OpenFlow, ForCES and PCE Protocol (step 4). The switch then forwards the packet to the appropriate port to send the packet to the receiver (step 5) [9].

## 8 Architecture Survey

Some of the selective architectures are discussed as follows:

1. Since presently optical equipment does not provide any support to OpenFlow and to control a wavelength switched optical network using OpenFlow protocol has not been investigated so far. Liu et al. presented a proof-of-concept to control a wavelength path in transparent optical network by using two different approaches for lightpath setup i.e., Sequential and delay approach and two different approaches for lightpath release i.e., active and passive approaches. To setup lightpath between sender and receiver, various optical nodes are interconnected to form optical network. In this paper, optical node comprising of OpenFlow switch and PXC. This combination is referred to as OpenFlow-enabled PXC (OF-PXC). In order to control this optical node through the OpenFlow protocol, they introduce "veths" concept and then this optical node is connected to the NOX controller as shown in Fig. 10. On the arrival of first packet of a new flow, OpenFlow switch encapsulates the packet and forwards it to the NOX controller. The controller analyses the packet to obtain source and destination IP address and accordingly assign route using k-shortcut path routing algorithm and assign wavelength using routing and wavelength assignment algorithm (RWA) on the basis of abstracted information. After this, NOX inserts new flow entry in the flow table of the OpenFlow switches, in response to this OpenFlow switch automatically generates transaction language-1 (TL-1) commands that instruct the PXC to cross-connect the corresponding ports using TCP interface to lay the lightpath.

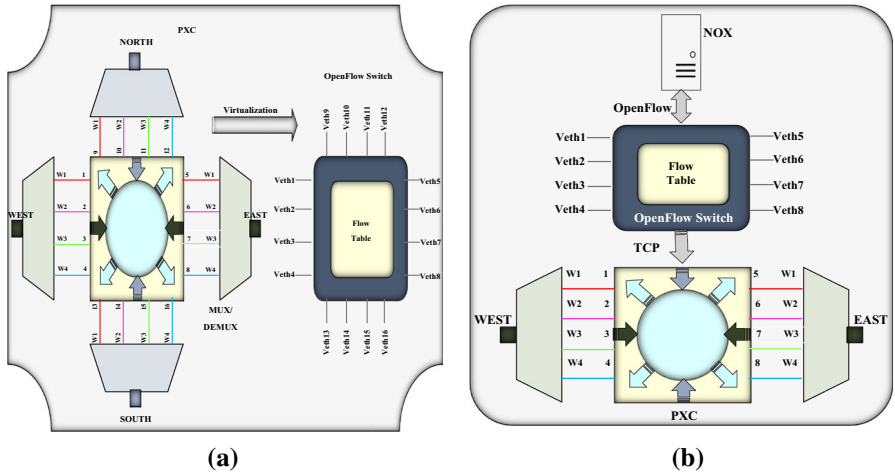

**Fig. 10** OpenFlow-based optical nodes [67]. **a** Virtualization of the physical interfaces of a PXC to virtual ethernet interfaces (*veths*) of an OpenFlow switch. **b** Architecture of an OF-PXC

*Sequential Approach* In this approach, OpenFlow protocol controls the NEs sequentially as shown in the Fig. 11. When packets of new flow arrive at the OF-R1, it encapsulates and forwards these packets to the NOX controller, and the NOX calculates the route using PCE and accordingly assigns wavelength to the optical network and update flow entries are inserted in the flow table of OF-R1, OF-PXC1, and OF-PXC2 respectively. OpenFlow switch of OF-PXC1 and OF-PXC2 sends TL-1 command to PXC for setup of cross-connects. Due to this, the lightpath is established between OF-R1, OF-PXC1 and OF-PXC2 successfully and light flow reached at OF-R2. OF-R2 checks the flow entry in flow table. If it does not find match, it sends the new flow packet to the NOX controller and NOX accordingly updates the flow entries of the OF-R2 and with this packet reaches to the destination.

*Delay Approach* In this approach, on the arrival of packet at the ingress OF-R1 node, matching is done, if does not found flow entry in the flow table of OF-R1, OF-R1 encapsulate and forwarded the packet to the NOX controller. NOX calculates the routing path between the source and the destination using PCE and accordingly assigns wavelength in the optical network. In delay approach after this, NOX adds flow entries in the flow table of the OF-PXC1 and OF-PXC2. Firstly, it establishes the light path in the optical network and after this, NOX deliberately adds a delay of few nano/milli second and then enters the flow entries in the ingress OF-R1 router. As the control of ingress node (OF-R1) is delayed due to this, this approach is called as delay approach as shown in the Fig. 12.

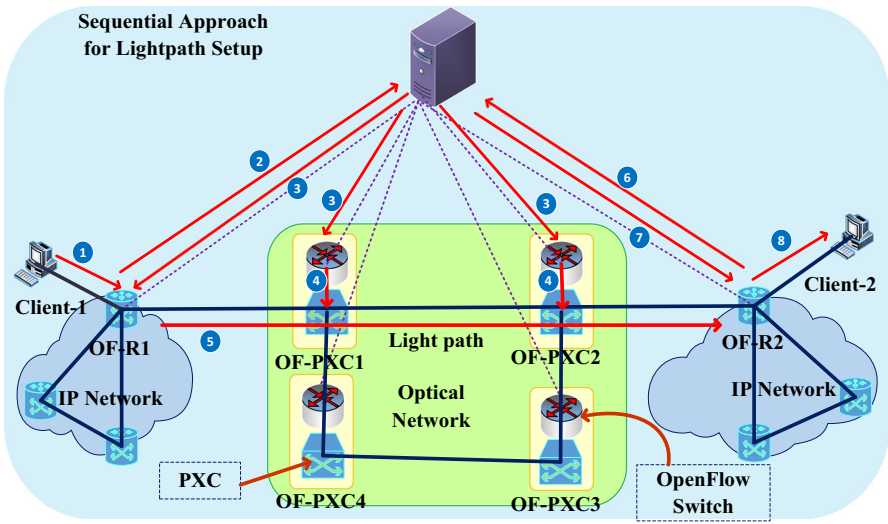

**Fig. 11** Sequential approach for lightpath setup [67]

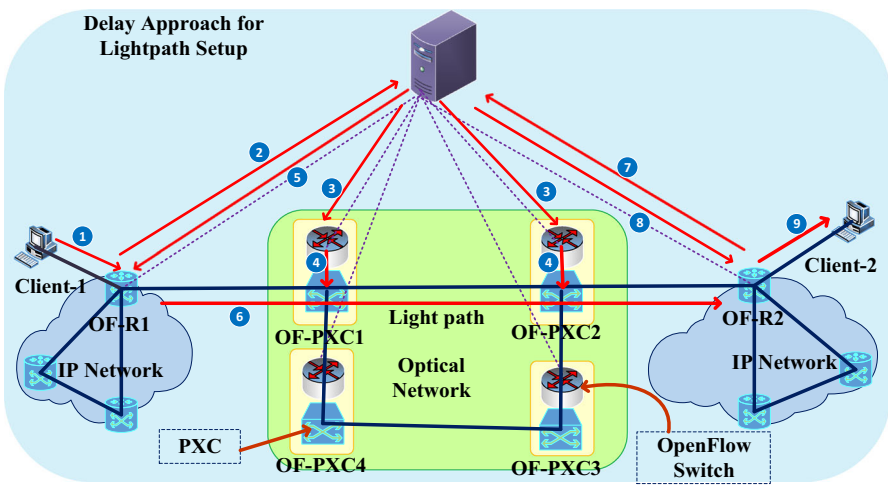

**Fig. 12** Delayed approach for lightpath setup [67]

*Active Approach* This approach is used when the amount of the data to be transferred is known well in advance, therefore the connection holding time can be predicted in the optical domain. NOX controller adds "hard timeout" field in the flow entry of the OpenFlow switch. This indicates the connection holding time. When the "hard time" is expired, flow entry in the OpenFlow switch is automatically deleted and cross-connection of the PXC is released and therefore, no further transmission of packets takes place.

*Passive Approach* In this approach, at the completion of packet transmission, the client sends packet to the ingress router/switch i.e., OF-R1 which consists of source and destination TCP/UDP port indicating the completion of flow transmission. After that the OF-R1 forwards this packet to the NOX controller as a first packet of new flow. NOX after analyzing, the destination TCP or UDP of these packets and predicts that it is not a first packet of new flow, but these packets indicate the transmission flow completion, therefore accordingly NOX controller deletes the flow entries from the flow table of OF-R1, OF-PXC1 and OF-PXC2. In response to this, OpenFlow switch sends TL-1 command to PXC for release of cross-connection.

Liu et al. [67] proposed an architecture, which consists of four PXC, connected in mesh topology, two IP router/switch i.e., ingress and egress nodes and two clients i.e., sender and receiver. Four different wavelengths are assigned to the optical link. Authors quantitatively analyses the network performance by considering various parameters, such as dynamic allocation of bandwidth by using RWA algorithm and light path setup and release latency in optical network. They observed that delay approach has an advantage as compared to sequential approach, as delay approach provides guaranteed successful end-to-end packet transmission without loss of any packet as light path in optical domain is well established before the arrival of the new flow packet at the ingress node. In passive approach lightpath released latency increases, as optical network complexity increases as compared to the active approach. With the help of OpenFlow protocol centralized controller has to inform more OpenFlow switches for the release of the cross-connection which increases signal processing latency.

2. Future internet is visualized to have characteristics that can have the potential to deliver packets globally using high performance network application like cloud computing, Big Data, Fog Computing, UHD video on demand etc., and bandwidth allocation depends on the traffic generation by these applications. When aggregated to transport over backbone/core network high-capacity Wave Division Multiplexing (WDM) circuit switching network is the only alternative. Siamak Azodolmolky et al. proposed an architecture that consists of unified control plane platform to integrate the electronic packets and optical networks for access, metro and core network segments. For this, it uses the services of OpenFlow protocols and GMPLS control plane to control and manage the software defined packets over the packet switching and circuit switching optical network as shown in Fig. 13. In this, authors consider the use-case of on demand UHD video content access and the corresponding timing diagram illustrates the various events that occur in a sequential way as shown in the

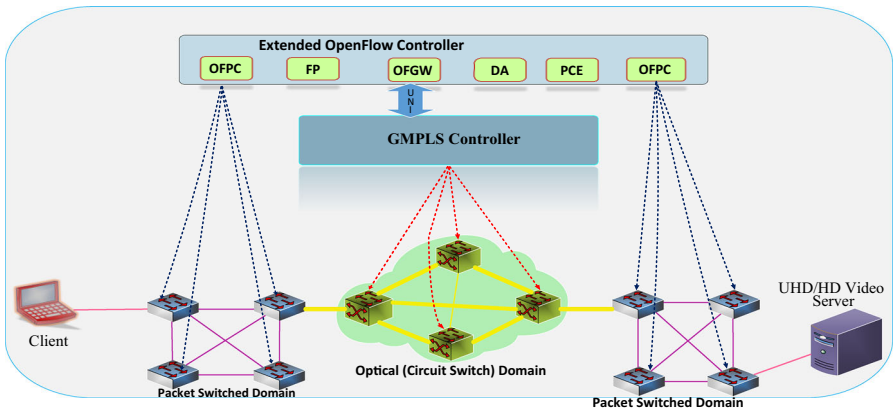

**Fig. 13** Integrated architecture of OpenFlow-GMPLS control plane [97]

Fig. 14. Client sends request for UHD video access to ingress OpenFlow switch, while the ingress OpenFlow switch treats it as first packet of a new flow. Ingress switch encapsulates and forwards packet to the extended OpenFlow controller. Controller processes this packet and identifies the endpoints or resolves the destination address and generates a request via OpenFlow gateway (OFGW) to GMPLS control plane using user network interface (UNI) for a new optical light path setup between the client and the server. After establishing a light path, GMPLS control plane acknowledges the extended OpenFlow controller. After this, controller updates their flow table of the ingress and egress OpenFlow switches and request is forwarded to the video server. In response to this server acknowledges to the client by providing the access to the UHD video contents.

*Experimental Setup and Results* Experimental network setup comprises two packet switches network and one circuit switching optical network. The flow of packets is controlled and managed by using OpenFlow protocols and GMPLS control plane services that is OpenFlow is used to separate the data path of the generic switching elements e.g. routers, switches, and APs from the control plane, whereas GMPLS control plane is used as a control for core optical network. In their work they consider parameter that is number of hops per optical flow versus average end-to-end flow setup time (s) to evaluate the performance of the unified control plane network and concludes that as number of hops per optical flow increases the average end-to-end flow setup time also increases [97].

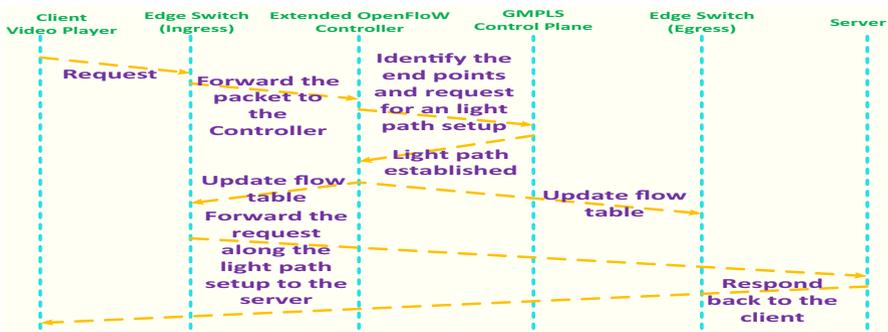

**Fig. 14** Timing diagram illustrating the various events occurs for the light path setup [97]

3. With the advancement of time optical technologies are being deployed to scale and flexible services, which are cloud, based and are encompassing network to new boundaries in SDN. The two key technologies such as SDN and flexible grid optical transport technology play a major role for the network operators to customaries their infrastructure which reduces the extra capital and operation cost while hosting new application. Channegowda et al. developed a unified control plane architecture approach as shown in Fig. 15, for multi-domain and multi-transport network based on SDN framework with OpenFlow. OpenFlow protocols are extended with a view to support fixed and flexible grids optical DWDM network along with multi-domain operation. With the help of OpenFlow protocol and GMPLS control plane and the networking devices, capabilities and constraints are abstracted and provided to the OpenFlow controller. By using this abstracted information, controller builds a technology and domain specific topology database. The information stored in the topology database is used by OpenFlow controller to facilitate the application of SDN related to PCE and virtual optical network provider (network slicer) to provide a path or network slice across varied technology and domain. Authors in their proposed architecture integrate multi-domain such as packet switches, fixed grid optical network, and flexible grid optical network and multi-transport technology, such as, electronic packets for campus/access/local and metro network and optical packets for backbone/core network for controlling and managing packet flow. In fixed grid domain, the network operators allocate fixed size optical spectrum ranging from 50 to 100 GHz for each channel spacing, whereas, in case of flexible grid optical domain, both channel spacing and channel bandwidth are variable and flexible.

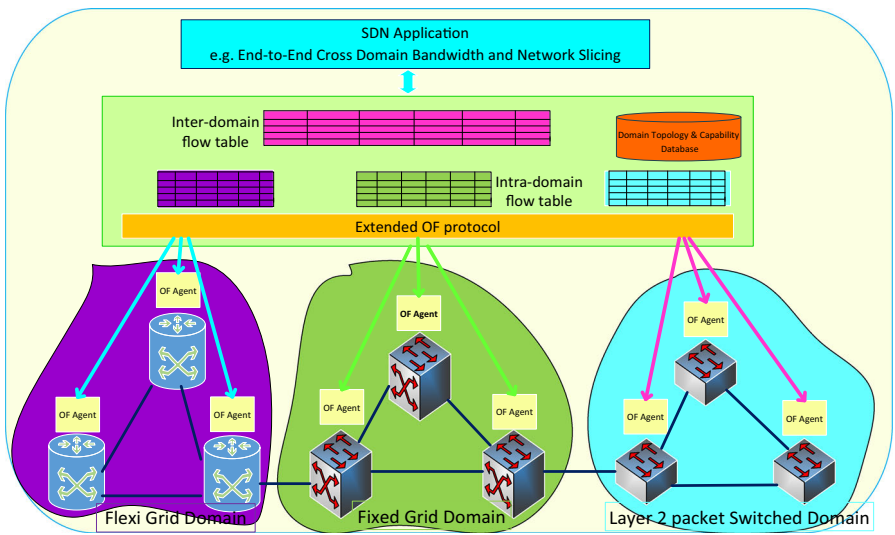

**Fig. 15** Architecture of multi-domain multi-technology control plane [12, 66]

In their proposed work they develop a bundle of algorithm and selection of the algorithm depends on domain (i.e., fixed grid or flexi-grid) e.g. if it's a flexible grid request, then the path is calculated using routing and spectrum assignment algorithm (RSA) and if it's fixed-grid domain then RWA is selected. The algorithm such as hop count shortest path routing is mostly used to trace a physical path for each single virtual link in order to reduce the number of NE involvement or domain. They also propose how new bandwidth is allocated in single/multiple or fixed/flexible domain. In fixed-grid domain, the first-fit wavelength assignment is used afterwards to allocate the required channels. In flexi-grid domain, among all the available spectrum slots, the one that can have the minimum residual spectrum after the spectrum assignment for the requested bandwidth is chosen.

They experimentally evaluate the performance of the proposed architecture, which is geographically distributed and comprises heterogeneous multi-domain, such as, flexible and fixed grid optical domain along with layer-2 packet switched domain. In their experimental setup, they use both hybrid i.e., GMPLS and OpenFlow agent and standalone OpenFlow approach, which includes OpenFlow agent on each NE. On the basis of above two approaches, they evaluate the performance by considering various parameters, such as, different path setup times, blocking rate of approaches, controller throughput performance and hardware setup time versus load. From this study they concluded that the hybrid approach is better than the standalone OpenFlow agent approach [12, 66].

4. In this paper, Guo et al. proposed a generic architecture as shown in Fig. 16 that supports various applications like DC, cloud computing and large-scale scientific computation, wherein, huge amount of data is transferred between end-systems, which are geographically distributed. In their work they introduce extended SDN controller, which is constructed by adding three application specific modules in it, like performance monitoring module, flow convergence module and rate control module. Control plane of the optical circuit switching network uses the services of GMPLS protocol to communicate with optical switches. Whereas, in case of packet switch network extended SDN controller uses the services of OpenFlow protocol to communicate with the OpenFlow switches. With the help of these protocols extended SDN controller abstract the network information (logical mapping of the network), so that dynamic allocation and optimized utilization of the bandwidth can be done with guaranteed transmission performance without packet loss.

In their experimental demonstration they evaluated the performance of the complete network by allocating different bandwidth to different services provided by the network at different flow rate. From the experimental result they observed that with their extended SDN controller bandwith utilization is improved or bandwidth wastage reduces, total/aggregate data transfer time reduces, therefore, latency and packet losses during the transmission are also reduced [98].

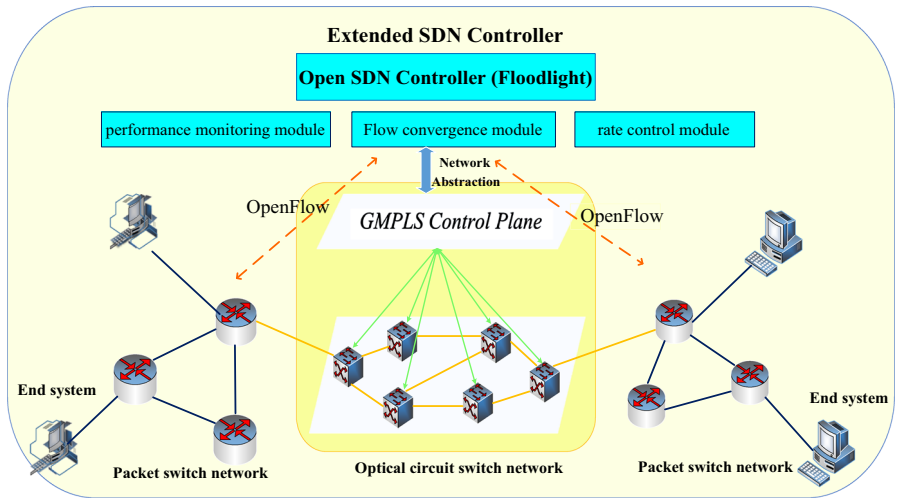

**Fig. 16** The architecture of extended SDN controller [98]

5. Applications like cloud computing, video gaming, UHD video streaming, live concerts, remote medical surgery and other applications are offered by DCs. These DCs are geographically distributed and connected via a network. Many decisions are made in the Application space without any concern of the underlying network. On the other hand, in order to achieve the optimization of application and network resource, cross stratum optimization (CSO) is proposed, which can enable a joint optimization of application and network resources.

   Yang et al. proposed centralized control architecture i.e., enhanced software defined network (eSDN) in place of elastic Grid (eGrid) optical network with a view to have migration of DC services by implementing a strategy known as transport aware cross stratum optimization (TA-CSO). eSDN can have the ability for CSO of application and eGrid optical network stratum resources and can also have the provision of adjusting elastic physical layer parameters e.g. bandwidth and modulation format. The distributed DC networks are connected among themselves with eGrid optical networks, which install network stratum resources and application respectively. Each stratum resources are software defined with OpenFlow and controlled by application controller (AC) and transport controller (TC) respectively in a unified manner as shown in Fig. 17. The proposed architecture consists of two controllers namely TC and AC. In this TC collects the information about the network status and accordingly constructs the global view of the entire network i.e., network topology graph and makes available this abstracted resource information to AC, whereas AC is responsible for controlling and discovery of the modulation format and spectrum bandwidth.

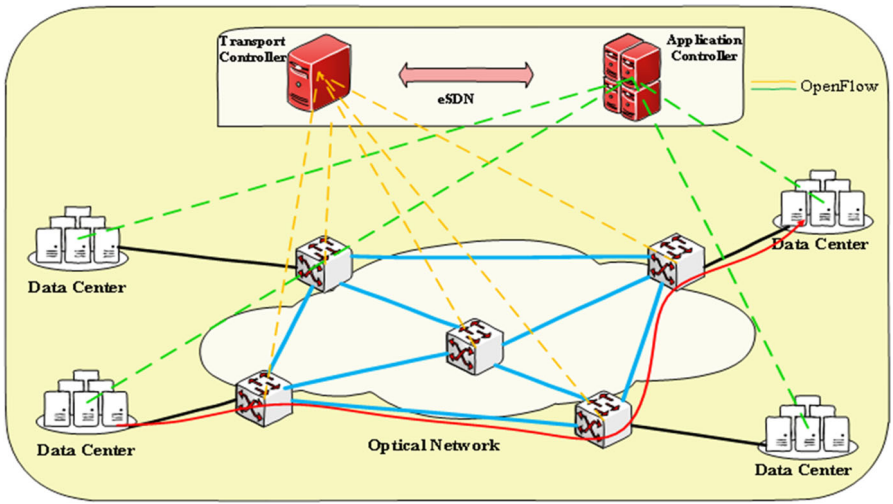

**Fig. 17** The architecture of OpenFlow-enabled SD-OTN (eSDN) comprise of AC and TC [99]

When request is arrived from the DC services, AC agrees to apply TA-CSO strategy of application and network resource information that is stored in internal database and directs the results to TC through application-transport interface (ATI). TC on getting service request from AC, calculates software defined path (SDP) from source-to-destination using extended OpenFlow Protocol. The authors evaluated the performance of the proposed architecture under dense traffic load situation and compared TA-CSO algorithm with individual CSO and physical layer adjustment strategies (PLA) in relations to resource occupation rate and blocking probability. They observed that when the network is heavily loaded the blocking probability decreases effectively by using TA-CSO as compared to CSO and PLA [99].

## 9 Software Defined Heterogeneous Network (SDHN) Architecture

Our proposed SDHN as a FN architecture is as shown in Fig. 18a, b comprises of controller, OpenFlow switch, packet switching network, optical circuit switching network, and base station/AP which provides seamless interfacing to N number of clients. In our proposed SDHN architecture the control plane contain centralized unified SDN controller that performs various functions such as route management, network visibility, orchestrates network overlays etc. and also communicate with data plane which further consist of heterogeneous network devices like Packet switching, optical switching and wireless devices as shown in Fig. 18a, b via a south-bound interface i.e., OpenFlow protocol and OpenFlow agent/GMPLS.

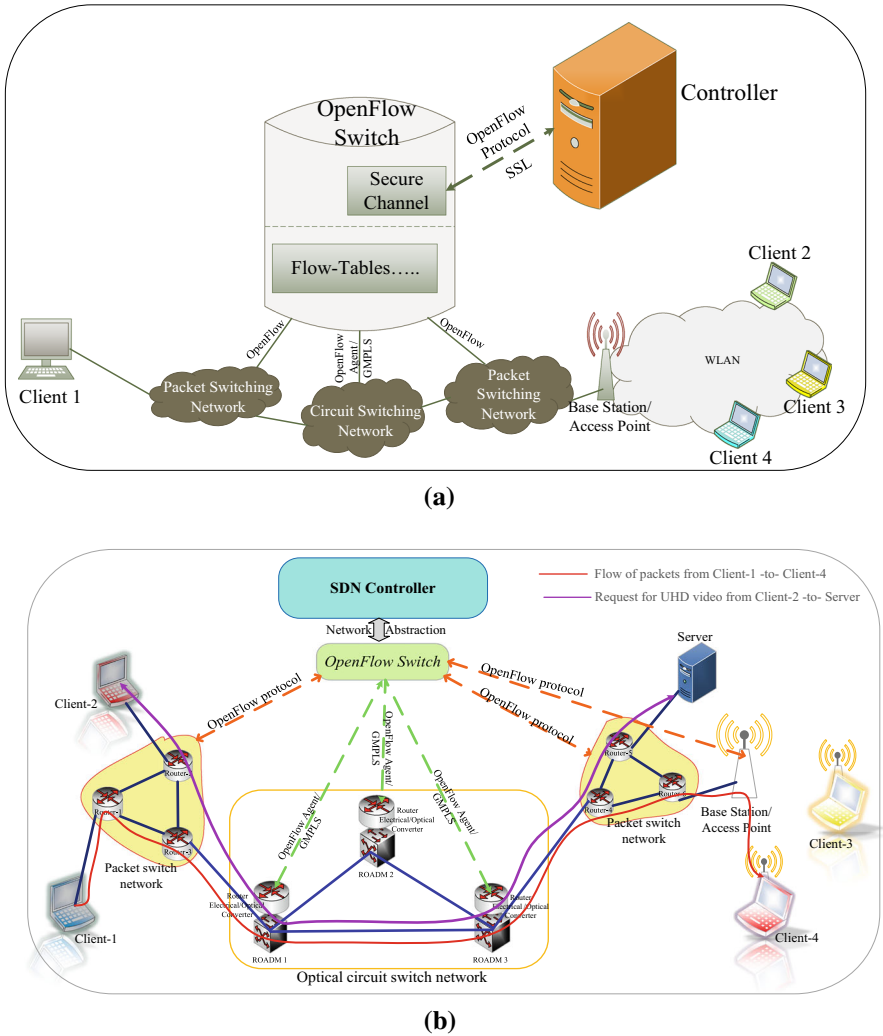

**Fig. 18 a** Block diagram of proposed SDHN, **b** the proposed architecture of the SDHN

When data is send from source to destination i.e., from Client-1-to-Client-4, the ingress router/switch i.e.Router-1first analyses the packet header and matching is done with the entries in flow table. If the flow table entry is matched with the header of the packets, then that particular entry is considered. Finally, the routers/switches in the data plane execute the action on the packets in accordance with the entry in the flow table i.e., forward packets to the destination. In case, if the packet header does not find match with the flow table entry of ingress router/switch i.e., Router-1 to forward these packets, then it encapsulate the first flow packet in PACKET_IN message as first byte of the new flow to the controller. On receiving PACKET_IN notification from the ingress router/switch, the controller calculates the route from Client-1 to Client-4 using PCE and sends PACKET_OUT message to the OpenFlow switch which holds the inter domain flow tables entries and accordingly update flow entries are reflected into the flow tables of routers/switches on the path in the network of data plane. Simultaneously, the controller sends OFPT_FLOW_MOD messages (OpenFlow messages mapping solution) or OFPT_CFLOW_MOD message (OpenFlow extension solution) to the OpenFlow agent/GMPLS in order to allocate wavelength to the optical network. On receiving this message from the controller OpenFlow agent/GMPLS translate it into appropriate TL1 commands and send it to the ROADM switches for creation of appropriate lightpath for packet flow. Finally the packet is received by the destination i.e., Client-4 via Router-1, Router-3, Router-ROADM-1, Router-ROADM-3, Router-4 and Router-6 respectively.

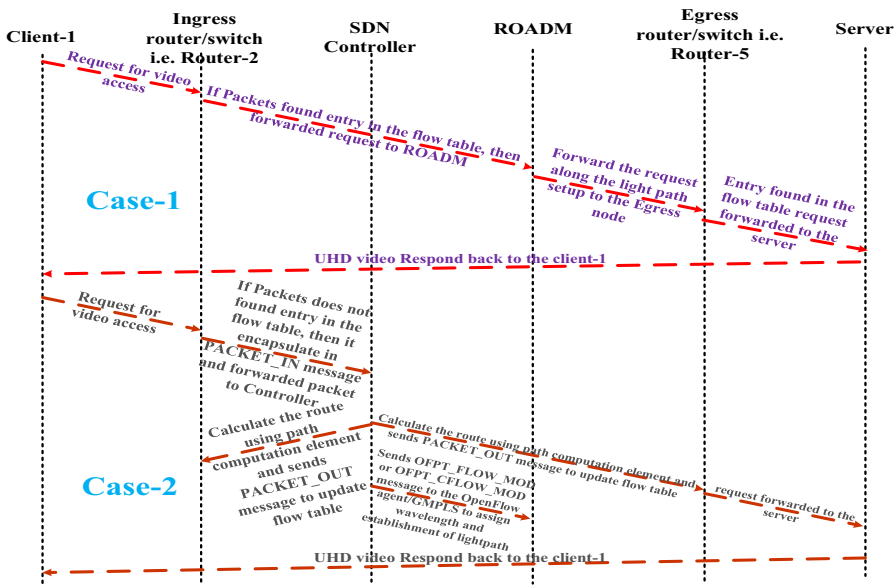

**Fig. 19** Illustrate various events that are initialed when client-2 sends request for on demand UHD video access to the server

If Client-2 sends request for on demand UHD video access to the server, the various events that occur in a sequential way are as shown in the Fig. 19.

## 10 Current Research and Standardization of SDN as NGN

The underlying paragraph will enlighten about SDN technology, which is advancing towards standardization and deployment as NGN:

In 2011, Deutsche Telekom, Google, Facebook, Verizon, Microsoft, and Yahoo established an ONF to endorse deployment of SDN and OpenFlow-based networks [65]. Due to release of OpenFlow-enable products and solutions from time to time by various leading vendors, both academia and industry has shown keen interest in developing software project and deploying of OpenFlow-based networks, as a well-organized ecosystem around OpenFlow [100]. With the advancement of time the each new release of OpenFlow version, its specifications are uninterruptedly growing with new features as mentioned in Fig. 8. During the progression of OpenFlow standardization several OpenFlow compliant switches and controllers came into existence [101]. The detail of the current available OpenFlow compliant switches is given in Table 6.

Current controller implementations compliant with OpenFlow standard is given in Table 5.

### 10.1 Ongoing Research Efforts

In most of the studies so for carried out experiments are laid in local area network (LAN). However, with the advancement of time wide area network (WAN) could also find its place. Das et al. [123] show that WAN could be implemented by deploying OpenFlow and the same is endorsed by further studies such as [124, 125].

Similar studies have also been carried in the field of mobility and wireless networks; it seems that the distributed control plane approach is an inefficient in managing limited resources such as spectrum, handover mechanisms, load balancing amid cells etc. Whereas, SDN-based approaches/methodologies make it more efficient, flexible, simpler to implement and easy to manage wireless networks like cellular and WLAN [126–131] via dynamic spectrum allocation [132], on-demand virtual access points (VAPs) creation [126, 132], well-organized handovers techniques [126, 130, 133], allocate efficiently base station resources management per client i.e., in long term evolution (LTE)/orthogonal frequency-division multiple access (OFDMA) resources are frequency and time slots [128, 129, 131] etc. and provides a platform, which helps in deployment of novel applications easily [126, 129, 134]. SDN could enables vital functions such as virtual network management and operation, network function virtualization (NFV) etc. that helps in development and deployment of huge capacity and gigantic connectivity of complex and powerful heterogeneous 5G wireless network [135, 136]. First example, OpenRoad can be viewed as Wireless version of OpenFlow to carry research in mobile networks. OpenRoads' architecture includes several wireless technologies such as WiMAX and WiFi [127, 137], deployment of the same in Stanford University is elucidated in [138]. For group communication on phones an infrastructure is proposed using PhoneNet as illustrated by Huang et al. [139]. The detail of the current research projects are given in Table 7.

### 10.2 Standardization Efforts

Recently, various efforts are being made to standardize the SDN-based network via SDOs/community consortia, the detail of which is given in Table 8.

Table 6  Currently available OpenFlow compliant commodity and its manufactures [3, 33, 65, 102]

| Category | References | Manufacturers/ vendors | Commodity model | Type | Version | Brief description |
| --- | --- | --- | --- | --- | --- | --- |
| Hardware switches | [103] | Arista | 7150 series | Switch | v1.0 | Integrating arista extensible operating system (EOS) natively with OpenStack. In heterogeneous network it enables the flexibility and freedom to implement both IP and OpenFlow |
|  | [104] | Brocade | NetIron CES 2000 series | Switch | v1.0 | Ethernet switches having high-performance ability to handle rising telecommunications bandwidth demands |
|  |  |  | MLX series | Router |  | It supports bigger bandwidth, and provide up to 128 ports of ten GbE |
|  | [65] | Cisco | Cisco 6500 series | Switch | v1.3 | It provides comprehensive network services in optimized campus core and distribution network |
|  |  | Dell | Dell Z9000 | Switch |  | Designed for high-density 10/40 GbE in a DC core network |
|  | [105] | Huawei | CX600 series | Router | v1.0 | Design to deploy at access and convergence layer of a carrier-class metropolitan area network (MAN) |
|  | [106] | Hewlett-Packet | HP procurve series-5400zl, and 8200zl | Chassis | v1.3 | Designed for deployment in Enterprise edge/small core and support for 1/10/100GbE ethernet |
|  | [107] | IBM | RackSwitch G8264 | Switch | v1.0 | First, 10 GbE switches that provides benefits of OpenFlow |
|  | [65] | Juniper | Juniper MX-series | Router | v1.0 | It provides high performance that enterprises, service providers and cloud operators required |
|  | [108] | NEC | NEC IP8800, PF5240, PF5820 | Switch | v1.0 | Hybrid technologies switch, with 48 ports of gigabit and four ports of 10 Gbps connectivity |
|  | [109] | NetFPGA | NetFPGA | Card | v1.0 | First, OpenFlow switches, implemented on open network hardware. Processing speed 1 or 10 Gbps. Port density is four ports and having medium flexibility |
|  | [110] | Plexxi | Plexxi Switch 2 | Switch | v1.0 | Designed to fulfill increasing demand for virtual and highly dynamic DC and cloud environments. It provides optical interfaces, enabling sending up to 480 Gbps full-duplex bandwidth using WDM |
|  | [111] | Pica8 | P-3290 | Switch | v1.0 | Hybrid ethernet/OpenFlow switches |

**Table 6** continued

| Category | References | Manufactures/vendors | Commodity model | Type | Version | Brief description |
|---|---|---|---|---|---|---|
| Software switches | [43, 112, 113] | Open community | Open vSwitch | Switch | v1.0–1.3 | Open source Software switch designed with an objective to implement switch platform in virtualized server environments using C/Python language |
| | [114] | Stanford | Pantou/OpenWRT | Switch | v1.0 | Implemented on general PC hardware. It turns AP/commercial wireless router into an OpenFlow enable switch. Processing speed low. Highly flexibility because of all features are software defined using C language |
| | [115] | Indigo | Indigo | Switch | v1.0 | Open source OpenFlow implemented on vendor's switch, with high processing speed i.e., 1 Gbps or above that has low flexibility using C language |
| | [116] | Ericsson, CPqD | Ofsoftswitch13 | Switch | v1.3 | OpenFlow 1.3 compatible user-space software switch implementation using C/C++ language |
| | [117] | Big switches | Switch light | Switch | v1.0 | Linux-based, thin switching software for physical and virtual switches which provide smooth migration path to SDN |
| | [118] | Yogesh Mundada | OpenFlowClick | Vrouter | v1.0 | OpenFlow element for Click, which allows hybrid packet and flow processing. |
| | [119, 120] | FlowForwarding | LINC | Switch | v1.2 and 1.3.2 | OpenFlow software switch written in Erlang. It provides sufficient elasticity and permits fast development and testing of new innovation of OpenFlow features |
| | [121] | Juniper2114 network | Contrail-vrouter | Vrouter | v1.0 | Data-plane functionality that permits a virtual interface to be allied with a VRF and liable for sending packets from one virtual to other VMs through a set of server-to-server tunnels |
| | [122] | Pica8 | XorPlus | Switch | v1.0 | Open source software, that runs on DC grade of switch platforms to provide high performance of switching/routing speed |

**Table 7** Current research project

| Site address | Project title | Name of agency | Organization | Overview |
|---|---|---|---|---|
| http://onrc.stanford.edu/projects.html | P4: high-level language for programming protocol-independent packet processors | ONRC | Stanford, Berkeley | P4 was proposed in association with Barefoot Networks, Google, Intel, Microsoft, and Princeton University as a high-level language for programming protocol-independent packet processors having three goals: (1) re-configurability in the field, (2) protocol independence, (3) target independence |
| | OpenRadio and software defined cellular wireless networking | ONRC | Stanford, Berkeley | At present, mobile wireless networks facing two conflicts: (1) the demand side, (2) the supply side. Moreover today, network is ill-equipped to handle future dense and dynamic infrastructure. A novel network architecture called OpenRadio was proposed as "software defined wireless network" |
| | SoftRAN | ONRC | Stanford, Berkeley | SoftRAN is software defined radio access networks. It enables a fresh look on the network architecture with two benefits: (1) each and every network service is considered as software application, (2) reduce capital and operation expenditure because of easy to development and deployment new network services |
| | Procel | ONRC | Stanford, Berkeley | Todays, mobile network have certain scaling issues because it treats traffic uniformly and due to this disproportionate allocation of network resources on data flow. Procel enables to rethink at mobile core transport network architecture and to eliminate this problem by categorizing data flow on the basis of their requirement and the value they deliver to the end-user, and accordingly allotting core network resources |
| | SDN for dense home network | ONRC | Stanford, Berkeley | As in metropolitan cities large number of population lives in multistory buildings, where dense WiFi networks are deployed with factory settings, leads to unsatisfactory resource utilization. Whereas, "Personal Network" leads to segregate the underlying physical NEs from logical view, which helps to fulfilling the end users' needs to customize and personalize their own network service |

**Table 7** continued

| Site address | Project title | Name of agency | Organization | Overview |
|---|---|---|---|---|
| | Sergeant, a distributed operating system (OS) for SDN | ONRC | Stanford, Berkeley | Sergeant is a distributed OS for SDN, proposes a new distributed scheduling algorithm permitting fine-grained parallelism |
| | SDN version controller | ONRC | Stanford, Berkeley | Propose distributed controller which keeps the records of flow table's changes and reserve histories of the events that cause these changes to the network. This further helps in debugging, analytics, and security forensics of the network applications |
| | Network virtualization | ONRC | Stanford, Berkeley | OpenVirteX has been developed in association with Open Networking Laboratory (ON.Lab), which provide NV podium that permits operators to create and manage virtual software defined networks (vSDNs) |
| http://pane.cs.brown.edu/ | Participatory networking | PANE | Brown University | Participatory networking is implemented by PANE, which is a prototype OpenFlow controller. In it Network Information Base maintains data about network topology for the distributed OpenFlow switches |
| http://projectbismark.net/ | Broadband internet service benchmark | BISmark | Georgia Tech, University of Napoli Federico II | BISmark is a research project working on open platform home broadband internet. In it performance of ISP is measured and traffic pattern is visualized using suitable devices while being at home. It is an collaborative research project between Georgia Tech. M-Lab, and Princeton University |
| http://incntre.iu.edu/SDNlab | – | InCNTRESDNLab | Indiana University | InCNTRE primarily provides education opportunities through SDN Interoperability Lab, wherein SDN technology having OpenFlow is being developed and adopted |
| http://www.ict-crowd.eu; | FP7 CROWD | FP7 | Mainly organizations in Europe | Providing networking architecture for MAC control and mobility management |

Table 8 Activities carried out by various SDOs that lead to the standardization of SDN as NGN

| References | SDOs | Brief description |
|---|---|---|
| [140–145] | Open Networking Foundation (ONF) | Originally, ONF was constituted to encourage the adaption of SDN by standardizing OpenFlow protocols in 2011, as an open standard to communication control decisions to data plane devices. ONF is organized in various working groups (WGs) they are |
| | | Architecture and Framework WG emphasis on SDN architecture and its architectural components |
| | | Interfaces WG concentration on data-controller plane interface (D-CPI) also called Southbound interface (SBI), provides interface between SDN controller and the underlying infrastructure under direct control and application-controller plane interface (A-CPI) also called northbound interface (NBI), provides interface between application and SDN controller |
| | | Extensibility WG responsible for development and deployment of extensions to OpenFlow protocol |
| | | Optical transport and wireless and mobile WG focus on specification and control capabilities mechanism for optical transport and wireless and mobile networks by mean of OpenFlow |
| | | Migration and market education WG emphasis on smooth transition from conventional to SDN-based network by means of OpenFlow and educate about SDN and OpenFlow technology by releasing white papers and solution briefs |
| [146–152] | Internet Engineering Task Force (IETF) | Network programmability concepts influence several Working Groups (WGs) of IETF and they are |
| | | ALTO WG emphasis on optimization of P2P traffic |
| | | ForCES WG standardize information exchange between the control and forwarding plane in a ForCES NEs |
| | | In an IP routed network Interface to the routing system (I2RS) WG focus on real time/event driven interaction with the routing system |
| | | PCE WG focus on PCE protocol, which is most commonly used protocol between control and physical layer |
| | | Source packet routing in networking (SPRING) WG focus on specification of a forwarding path at the source of traffic |
| [153–156] | Internet Research Task Force (IRTF) | IRTF has proposed a SDN Research Group (SDNRG) that examines/identifies various SDN approaches and their deployed in the nearby future as well as recognize various future research challenges |

Table 8 continued

| References | SDOs | Brief description |
|---|---|---|
| [157–163] | International Telecommunications Union's Telecommunication (ITU-T) | In ITU-T various study groups (SGs) started to develop recommendation for SDN and a Joint Coordination Activity on SDN (JCA-SDN) has been constituted to coordinate the SDN standardization work. The various SGs are |
| | | SG 11 focuses on protocols and broadband access network signalling requirement using SDN technologies |
| | | SG 13 focuses on architecture and functional requirement of FNs |
| | | SG 15 focuses on architecture to provide support to transport network control plane of the SDN |
| | | SG 17 focuses on the security services and architectural aspects of security using SDN |
| [164] | Broadband Forum (BBF) | BBF works on SDN technology via service innovation and market requirements (SIMR) WG aiming to release recommendations that provide support to SDN technology in multi-service broadband networks |
| [165] | Metro Ethernet Forum (MEF) | MEF emphasis on SDN technology via the third network WG aims on service orchestration network and NFV environments |
| [102] | Institute of Electrical and Electronics Engineers (IEEE) | For both wired and wireless technologies, IEEE 802 LAN/MAN Standard Committee started some activities for standardizing SDN capabilities on access network via P802.1CF project |
| [166] | Optical Interworking Forum (OIF) | OIF via carrier WG released set of recommendation for transport SDN |
| [167] | Alliance for Telecommunication Industry Solution (ATIS) | ATIS constituted various focus group (FG) for investigating operational problems and opportunities related to the programmable network infrastructure |
| [168–173] | European Telecommunication Standards Institute (ETSI) | ETSI dedicated to Networking function virtualization via recently defined Industry Specification Group (NFV ISG) focuses on innovation inside network are also being done through automation/programmability and by considering SDN concepts as complementary |
| [174–177] | Distributed Management Task Force (DMTF) | DMTF established in 1992, in collaboration with various companies to development, validation and promotion of infrastructure management standards and is also responsible for integrating and management of diverse traditional and emerging technologies including cloud, virtualization, network and infrastructure |

## 11 Conclusion

With the advancement of time, manifold increase in networking traffic may be inevitable and it may not be feasible to provide efficient services with existing technology. Ever improvement in technology is a continuing process. In this survey paper, SDN as advancement over the conventional network, wherein control plane has been separated from the data plane is highlighted, which helps in increasing scalability, reliability and network performance. The journey of programmable network from its infancy to recent development spread over about last 20 years has been presented with brief description and contribution of authors. SDN architecture is also specifically discussed in all its facets of operation (technology), including OpenFlow standards/protocols in relation to interfacing NEs. As per one of the study, it has been observed that optical devices do not support OpenFlow standards/protocol, for which alternative approaches have been sought, that includes veths, OpenFlow agent, GMPLS and hybrid technique. Further, critical survey has been carried out on selective SDN architecture on the basis of services provided by the architecture, techniques used to interface the network element and various approaches used for optimum utilization of the underlying infrastructure and resources.

Our proposed SDHN as FN architecture comprises centralized unified SDN controller that performs various functions such as route management, network visibility, orchestrates network overlays etc. and also communicates with data plane, which further consist of heterogeneous network devices like Packet switching, optical switching and wireless devices. Besides, the technology involved in SDN has been duly elaborated through some of the current running projects indicating that how this technology moves towards standardization.

It may further be pointed out that Future Internet architecture will have to be based on *infrastructure as a service* (IaaS) commercial model that segregate the contribution of the existing *internet service providers* (ISPs) into twin novel roles, first related to *infrastructure provider* (InP), which deploys and maintains the NE and second deals with *service provider* (SP), which provides end-to-end service by deploying network protocols. Whereas, SDN via NV splits the roles of the SP into three key players: first, the *virtual network provider* (VNP) gathers virtual resources from single or multiple InPs, Second the *virtual network operator* (VNO) installs, manages and operates the VN as per the requirements of the SP, and third the *SP* that offers customized services using the VNs [178, 179]. From this, it may be concluded that SDN is one the most promising technology that permits the network administrators and service providers to customize their infrastructure dynamically based on the application requirements; so that capital expenditure and operation cost can be minimized by optimizing utilization of the underlying infrastructure and the resource. Thus, the SDN: architecture for NGN is to stay for undefined longer period, whose applications and implementations have yet to be fully exploited for use on wider scale on global basis.

**Sanjeev Singh** presently pursuing Ph.D. in E&C Engineering at SMVD University, Katra, Jammu and Kashmir, India. His research interest includes the Analysis of Opto-geometric Parameter's effects on Gain and Noise Figure of EDFA for WDM application. Currently, he is doing his research work on Software Defined Infrastructure based Network: Technology and its Optimization for optimize utilization of underlying infrastructure and resources of the NGN. Mr. Singh is a member of IEEE and IACSIT.

**Rakesh Kumar Jha (S'10, M'13)** is currently an Assistant Professor in School of Electronics and Communication Engineering, Shri Mata Vaishno Devi University, Katra, Jammu and Kashmir, India. He has received young scientist author award by ITU in Dec 2010. He has received APAN fellowship in 2011 and 2012, and student travel grant from COMSNET 2012. He is a member of IEEE, GISFI, and SIAM, International Association of Engineers (IAENG) and Advance Computing and Communication Society (ACCS).